\documentclass[10pt,twocolumn]{article}
\usepackage[top=1.8cm, bottom=1.8cm, left=1.8cm, right=1.8cm]{geometry}
\usepackage{times} 
\usepackage{titlesec} 
\titlespacing*{\section}{0pt}{*2}{2pt} 
\titlespacing{\subsection}{0pt}{*2}{2pt}

\usepackage[hyphens]{url} 
\usepackage{graphicx}
\def\UrlFont{\rm} 
\usepackage{graphicx} 
\usepackage{url}                                %
\usepackage{array,multirow,graphicx,adjustbox}  %
\usepackage{booktabs}                           %
\usepackage[utf8]{inputenc}
\usepackage[ruled,algosection,noend,linesnumbered]{algorithm2e}
\usepackage{float}
\usepackage{paralist}
\usepackage{amsmath}
\usepackage{amsfonts}
\usepackage{xcolor}
\usepackage{csquotes} 
\usepackage{tabularx}
\usepackage{makecell}
\usepackage{pbox}
\usepackage[keeplastbox]{flushend}
\usepackage{subfigure}
\usepackage[small,bf]{caption}
\usepackage[hang,flushmargin]{footmisc}
\usepackage{xspace}
\usepackage{multicol, lipsum}
\usepackage[T1]{fontenc}

\usepackage[bookmarks=true, bookmarksnumbered=true, colorlinks=true, linkcolor=linkcol, citecolor=citecol, urlcolor=urlcol, hypertexnames=true]{hyperref}

\definecolor{darkgreen}{RGB}{0, 100, 0}
\definecolor{linkcol}{rgb}{0.3,0,0}
\definecolor{citecol}{rgb}{0.3,0,0}
\definecolor{urlcol}{rgb}{0.3,0,0}

\makeatletter
\def\url@leostyle{%
  \@ifundefined{selectfont}{\def\UrlFont{\small}}%
  {\def\UrlFont{}}%
}
\makeatother
\newcommand{\descr}[1]{\smallskip\noindent\textbf{#1}}

\let\OLDthebibliography\thebibliography
\renewcommand\thebibliography[1]{
  \OLDthebibliography{#1}
  \setlength{\parskip}{0pt}
  \setlength{\itemsep}{1pt plus 0.2ex}
}

\title{\bf ``I Won the Election!'': An Empirical Analysis of Soft Moderation Interventions on Twitter\thanks{This paper is appearing in the 15th AAAI Conference on Web and Social Media (ICWSM 2021). Please cite the ICWSM version.}}

\author{Savvas Zannettou\\[0.5ex]
\normalsize{Max Planck Institute for Informatics}\\
\normalsize szannett@mpi-inf.mpg.de\vspace*{-0.2cm}}

\date{}

\begin{document}
\maketitle

\begin{abstract}
Over the past few years, there is a heated debate and serious public concerns regarding online content moderation, censorship, and the principle of free speech on the Web.
To ease these concerns, social media platforms like Twitter and Facebook refined their content moderation systems to support soft moderation interventions. 
Soft moderation interventions refer to warning labels attached to potentially questionable or harmful content to inform other users about the content and its nature while the content remains accessible, hence alleviating concerns related to censorship and free speech.

In this work, we perform one of the first empirical studies on soft moderation interventions on Twitter. 
Using a mixed-methods approach, we study the users who share tweets with warning labels on Twitter and their political leaning, the engagement that these tweets receive, and how users interact with tweets that have warning labels.
Among other things, we find that 72\% of the tweets with warning labels are shared by Republicans, while only 11\% are shared by Democrats.
By analyzing content engagement, we find that tweets with warning labels had more engagement compared to tweets without warning labels.
Also, we qualitatively analyze how users interact with content that has warning labels finding that the most popular interactions are related to further debunking false claims, mocking the author or content of the disputed tweet, and further reinforcing or resharing false claims.
Finally, we describe concrete examples of inconsistencies, such as warning labels that are incorrectly added or warning labels that are not added on tweets despite sharing questionable and potentially harmful information.
\end{abstract}

\section{Introduction} 

Social media platforms like Twitter and Facebook are under pressure from the public to address issues related to the spread of harmful content like hate speech~\cite{twitter_hate_speech2016} and online misinformation~\cite{facebook_disinformation2016}, particularly during major events like elections.
To ease the public's concerns and mitigate the effects of these important issues, platforms are continuously refining their guidelines and improving their content moderation systems~\cite{twiter_updates}.

Designing and implementing an ideal content moderation system is not straightforward as there are many challenges and aspects to be considered~\cite{gillespie2018custodians}.
First, content moderation should be performed in a timely manner to ensure that harmful content is removed fast and only a small number of users are exposed to harmful content.
This is a tough challenge given the scale of modern social media platforms like Twitter and Facebook.
Second, content moderation should be consistent and fair across users.
Finally, content moderation should ensure that it is in accordance with basic principles of our society like the freedom of speech.

\begin{figure}[t!]
\centering
\includegraphics[width=0.9\columnwidth]{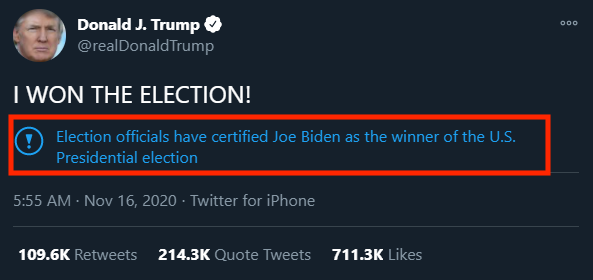}
\caption{An example of a soft moderation intervention on Twitter.}
\label{fig:example-softmod}
\end{figure}

To ease concerns related to freedom of speech and censorship, recently, Facebook and Twitter introduced a new feature in their content moderation systems; a type of soft moderation intervention that attaches warning labels and relevant information to content that is questionable or potentially harmful or misleading~\cite{twitter_softmod,facebook_softmod}.
An example of a soft moderation intervention is depicted in Fig.~\ref{fig:example-softmod}, where Twitter moderators attached a warning label to a tweet from President Trump related to the outcome of the 2020 US elections.
These warning labels are designed to ``correct'' the tweet's content and provide necessary related information while ensuring that the freedom of speech principle is not violated. 

Previous work investigated how users perceive these warning labels~\cite{mena2020cleaning,geeng2020social,saltz2020encounters,seo2019trust}, assessing their effectiveness, how their design can affect their effectiveness~\cite{bode2015related,kaiser2020adapting,moravec2020appealing}, and possible unintended consequences from the use of warning labels~\cite{pennycook2020implied,nyhan2010corrections}.
Despite this rich research work, the majority of it investigates these warning labels in artificial environments either through interviews, surveys, or crowdsourcing studies.
While these studies are useful and important, they do not consider platform-specific affordances such as user interactions with posts that have warning labels (e.g., retweets, likes, etc.)  
As a research community, we lack empirical evidence to understand how these warning labels are used on social media platforms like Twitter and how users interact and engage with them.

In this work, we aim to bridge this research gap by performing an empirical analysis of soft moderation interventions on Twitter. We focus on answering the following research questions:
\begin{itemize}
    \item \textbf{RQ1:} What are the various warning labels used on Twitter during the 2020 US elections and what kind of users have their tweets flagged more frequently? Are there differences across political leanings?
    \item \textbf{RQ2:} Is the engagement of content that includes warning labels significantly different from the content without warning labels?
    \item \textbf{RQ3:} How do users on Twitter interact with content that includes warning labels?
\end{itemize}

To answer these research questions, we collect a dataset of tweets shared between March 2020 and December 2020, including soft moderation interventions (i.e., warning labels). 
To do this, we use Twitter's API and collect the timeline of popular verified users. 
We mainly focus on verified users as they usually have a large audience and their content can receive considerable engagement. 
Overall, we collect a set of 18K tweets that either had warning labels on them or quote a tweet that had a warning label.
These tweets were shared from 8.1K users between March 2020 and December 2020. 
Due to our data collection method, 96\% of the tweets in our dataset were shared between November 2020 and December 2020, hence our dataset is related to the 2020 US elections.
Then, we follow a mixed-methods approach to analyze the engagement of tweets with warning labels and the users that share them (quantitative analysis), as well as how users interact with tweets and warning labels (qualitative analysis).

\descr{Findings.} Our main findings are:

\begin{compactenum}
    \item We find that 72.8\% of the tweets that include warning labels were shared by Republicans, while only 11.6\% of the tweets were shared by Democrats. 
    This likely indicates that during the 2020 US elections period, Republicans tended to disseminate more questionable or potentially harmful information (e.g., claims about election fraud) that is eventually flagged by Twitter (\textbf{RQ1}). 
    \item By analyzing the engagement of tweets, we find that tweets that have warning labels receive more engagement compared to tweets without warning labels. Also, by looking into the users that have increased engagement in tweets with warning labels, we find that most of the users that have high engagement, in general, have increased engagement on tweets with warning labels as well (\textbf{RQ2}). 
    Note that our analysis does not consider \emph{when} the warning label was placed and the engagement before the addition of the warning label. 
    Therefore, it is likely that the increased engagement on tweets with warning labels is because Twitter is prioritizing high engagement content within their soft moderation systems~\cite{twitter_softmod}.
    In other words, it is likely that the warning label was placed because the tweet already received high engagement.
    \item Our qualitative analysis indicates that a lot of users interact with content that has warning labels by further debunking false claims, mocking or sharing emotions about the author/content of the questionable tweet, or by reinforcing the false claims that are included in tweets with warning labels. Also, we shed light on some of the challenges and issues that exist when designing and developing large-scale soft moderation intervention systems. We find instances where the warning labels were incorrectly added (e.g., see Fig.~\ref{fig:mislabel_example}) and cases where the moderation system is inconsistent (i.e., content should be flagged, but it is not).
    Some of these cases are likely due to the dissemination of similar information across different languages (e.g., see Fig.~\ref{fig:language_example}) and across various formats of information like text and videos (\textbf{RQ3}).  
\end{compactenum}

\descr{Contributions.} The contributions of this work are three-fold. 
First, to the best of our knowledge, we perform one of the first characterizations of soft moderation interventions based on empirical data from Twitter.
Also, we make our dataset publicly available~\cite{dataset}, hence assisting the research community in conducting further studies on soft moderation interventions based on empirical data.
Second, our quantitative analysis sheds light on the engagement that content with soft moderation interventions receive on Twitter.
This analysis encapsulates engagement from real users interacting with timely content on Twitter, hence it complements and strengths the findings from studies undertaken in controlled experiments (e.g., via surveys).
Finally, our qualitative analysis highlights how users interact with content that includes warning labels, and it helps us understand some of the real-world challenges that exist when designing soft moderation intervention systems.

\section{Background \& Related Work}

Moderation interventions on social media platforms can be applied on various levels.
First, some interventions are applied at the post level (e.g., post-removal). 
Second, some interventions exist on the user level~\cite{myers2018censored,merrer2020setting} like user bans or shadow banning (i.e., limiting the visibility of their activity).
Finally, community-wide moderation interventions exist where platforms moderate specific sub-communities within their platforms (e.g., banning Facebook groups or subreddits)~\cite{chandrasekharan2020quarantined,chandrasekharan2017you,newell2016user,ribeiro2020does,saleem2018aftermath}. 

For each of the above-mentioned levels, there are two different types of interventions: \emph{hard} and \emph{soft} interventions.
Hard moderation interventions refer to moderation actions that remove content or entities from social media platforms (posts, users, or communities).
On the other hand, soft moderation interventions do not remove any content, and they aim to inform other users about potential issues with the content (e.g., by adding warning labels) or limiting the visibility of questionable content (shadow banning) or imposing restrictions on the ability of other users to engage with the content (i.e., users are unable to reply, re-share, etc.).
Below, we review relevant previous work that studies post-level soft moderation interventions as they are the most relevant to our work.

\subsection{Post-level Soft Moderation Interventions}

A rich body of previous work investigates soft moderation interventions mainly through interviews, surveys, and crowdsourcing studies.
Specifically, Mena~\cite{mena2020cleaning} performs an experiment using Amazon Mechanical Turk (AMT) workers to understand user perceptions on content that includes warning labels. By recruiting Facebook users and performing crowdsourcing studies, they find that the warning label had a significant effect on users' sharing intentions; that is, participants were less willing to share content with warning labels.
Geeng et al.~\cite{geeng2020social} focus on warning labels added on Twitter, Facebook, and Instagram related to COVID-19 misinformation. Through surveys, they find that users have a positive attitude towards warning labels. However, they highlight that users verify misinformation through other means and search on the Web for relevant information.
Saltz et al.~\cite{saltz2020encounters} focus on warning labels added on visual misinformation related to COVID-19. By conducting in-depth interviews, they find that participants had different opinions regarding warning labels, with many participants perceiving them as politically-biased and an act of censorship from the platforms.

Kaiser et al.~\cite{kaiser2020adapting} use methods from information security to evaluate the effectiveness and the design of warning labels.
Through controlled experiments, they find that despite the existence of warning labels, users seek information via other means, thus confirming the findings from~\cite{geeng2020social}. Also, by performing crowdsourcing studies and asking users about 8 warning label designs, they conclude that users' information-seeking behavior is significantly affected by the design of the warning label. 
Seo et al.~\cite{seo2019trust} investigate user perceptions when they are exposed to fact-checking and machine learning-generated warning labels. Through experiments on AMT, they find that users trust more fact-checking warning labels than machine learning-generated ones. 
Moravec et al.~\cite{moravec2020appealing} highlight that the design of warning labels (i.e., how warnings are presented to users) can change their effectiveness. Also, they emphasize that clearly explaining the warning labels to users can lead to increased effectiveness.
Bode et al.~\cite{bode2015related} study the related stories functionality on Facebook as a means to detect or debunk misinformation. By conducting surveys, they find that when related stories debunk a misinformation story, it significantly reduces the participants' misperceptions (beliefs that are not supported by evidence or expert opinion~\cite{nyhan2010corrections}).

Other previous work demonstrates some unintentional consequences from the use of warning labels.
Specifically, Pennycook et al.~\cite{pennycook2020implied} conduct AMT studies and they show an \emph{implied truth} effect, where posts that include misinformation and are not accompanied with a warning label are considered credible.
Also, Nyhan and Reifler~\cite{nyhan2010corrections} conduct controlled experiments to assess the effectiveness of warning labels to political false information.
They highlight a \emph{backfire effect}, where participants strengthen their support to false political stories after seeing the warning label that includes a correction. 
Follow-up research on the backfire effect~\cite{wood2019elusive} was unable to find instances of it, while another work~\cite{swire2020searching} suggests that this effect is highly dependent on context.
Finally, Pennycook et al.~\cite{pennycook2018prior} emphasize the existence of the \emph{illusory truth effect} where users tend to believe false information after getting exposed to it multiple times or for an extended time period, even though the false information is accompanied by a warning label. 

\descr{Remarks.} Previous work investigated soft moderation interventions in artificial testing environments like interviews, surveys, and crowdsourcing studies.
This previous work is essential as it helps us understand how people intend to interact and engage with content that includes warning labels or corrections.
However, they do not capture platform-specific peculiarities, and they do not adequately capture how people interact and engage with warning labels in realistic scenarios (e.g., when reading a tweet from the US President during the US elections period).
In this work, we address these limitations by performing, to the best of our knowledge, one of the first empirical analysis of soft moderation interventions on Twitter.

\section{Dataset} \label{sec:dataset}

We start our data collection on Twitter and, in particular, on verified users, which are users who have an ``especially large audience or are notable in government, news, entertainment, or another designated category.''\footnote{The definition is obtained from the Twitter website.}
We mainly focus on verified users as they usually have a large audience and can have a substantial impact on online discussions, hence moderating content from these users is important.

We collect the dataset of Twitter verified users from Pushshift~\cite{pushshift_verified}.
The dataset includes Twitter account metadata for 351,655 verified users.
Then, for each user, we use Twitter's API to obtain recent tweets/retweets shared by these users (i.e., their timeline). 
We also collect soft moderation-specific metadata for each tweet: these include whether a tweet is accompanied by a warning label and relevant metadata (e.g., label text, landing URL, etc.). 
Note that we only collect warning labels placed below the tweets and not any other types of interventions like tweets placed below an interstitial like ``This tweet may include sensitive content''~\cite{twitter_notices}.
Due to the rate-limiting of the Twitter API, we tried to collect activity only from the top 170,506 users based on the number of their followers (corresponding to 48.4\% of all the Twitter verified accounts in the Pushshift dataset).
We managed to collect data for 168,126 users, as the rest were either deleted, suspended, or accounts set to private.
Our dataset collection process was conducted between December 7, 2020 and December 31, 2020.
Overall, we collect 79,361,081 tweets shared during 2020 from 168,126 users.

Next, we select all tweets with soft moderation interventions (i.e., warning labels) from our dataset; we find 29,232 tweets from 9,334 verified users. 
This dataset also includes retweets of tweets with warning labels and tweets that quote a tweet with a warning label.
Due to this, we rehydrate, using the Twitter API, all quoted and retweeted tweets that had a warning label; we get an additional 3,106 tweets from 1,888 users.
Note that this procedure resulted in the acquisition of tweets from unverified users.
This is because verified users in our dataset retweeted or quoted tweets from unverified users.
Given that this content appears on the followers of verified users, we keep in our dataset tweets from unverified users.

After excluding all retweets, our final dataset consists of 18,765 tweets that include warning labels (either on the tweet itself or on referenced tweets like quoted tweets) from 8,142 users (see Table~\ref{tab:dataset}).
We split our dataset into two parts; 
1)~tweets that have warning labels attached to them (first row in Table~\ref{tab:dataset}); 
and 2)~tweets that quote other tweets and any (or both) of the tweets have warning labels (see second-fourth row in Table~\ref{tab:dataset}). Note that there is overlap on the tweets with warning labels and quoted tweets for the cases where the warning label was on the comment above.
For the remainder of this paper, we call the first part of our dataset \emph{tweets with warning labels} and the second part of our dataset \emph{quoted tweets}.
Note that due to our data collection approach, a large percentage of the tweets in our dataset (96\%) are shared between November 1, 2020 and December 30, 2020, hence our dataset is centered around the 2020 US elections and has a strong political nature.

\begin{table}[t]
\centering
\resizebox{\columnwidth}{!}{%
\begin{tabular}{@{}lrr@{}}
\toprule
                                                                                      & \multicolumn{1}{l}{\textbf{\#Tweets}} & \multicolumn{1}{l}{\textbf{\#Users}} \\ \midrule
\textbf{Tweets with warning labels (e.g., Fig.~\ref{fig:example-softmod})}                                                         & 2,244                                 & 853                                  \\ \midrule
\textbf{\begin{tabular}[c]{@{}l@{}}Quoted -  Warning on quoted (e.g., Fig.~\ref{fig:trump_mock})\end{tabular}}  & 16,571                                & 7,651                                \\
\textbf{\begin{tabular}[c]{@{}l@{}}Quoted - Warning on comment (e.g., Fig.~\ref{fig:language_example})\end{tabular}} & 219                                   & 98                                   \\
\textbf{\begin{tabular}[c]{@{}l@{}}Quoted - Warning on both (e.g., Fig.~\ref{fig:mislabel_example})\end{tabular}}          & 50                                    & 30                                   \\ \midrule
\textbf{Total}                                                                        & 18,765                                & 8,142                                \\ \bottomrule
\end{tabular}%
}
\caption{Overview of our dataset.}
\label{tab:dataset}
\end{table}

\descr{Ethical considerations and data availability.} We emphasize that we collect and work entirely with publicly available data as we do not collect any data from users who have a private account.
Overall, we follow standard ethical research standards~\cite{rivers2014ethical} like refraining from tracking users across sites.
To help advance empirical research related to soft moderation interventions on Twitter, we make publicly available~\cite{dataset} the tweet IDs and their corresponding warning labels.

\section{RQ1: Warning Labels and Users}
\label{sec:general_characterization}

In this section, we analyze the different types of warning labels and how they appear over time on Twitter.
Also, we perform a user-based analysis on users who shared tweets with warning labels or quoted tweets, aiming to uncover differences across users that have opposing political leaning.

\begin{table}[t]
\centering
\resizebox{\columnwidth}{!}{%
\begin{tabular}{llr}
\hline
\textbf{\#} & 
\textbf{Warning label}                                                                                       & \multicolumn{1}{c}{\textbf{Tweets}} \\ \hline
1 & This claim about election fraud is disputed                                                                                            & 1,305 (58.1\%)                      \\ \hline
2 & Learn about US 2020 election security efforts                                                                                          & 271 (12.1\%)                        \\ \hline
3 & Manipulated media                                                                                                                      & 196 (8.7\%)                         \\ \hline
4 & Learn how voting by mail is safe and secure                                                                                            & 132 (5.8\%)                         \\ \hline
5 & \begin{tabular}[c]{@{}l@{}}Official sources may not have called \\ the race when this was Tweeted\end{tabular}                         & 101 (4.5\%)                         \\ \hline
6 & Multiple sources called this election differently                                                                                      & 96 (4.2\%)                          \\ \hline
7 & \begin{tabular}[c]{@{}l@{}}Election officials have certified Joe Biden\\  as the winner of the U.S. Presidential election\end{tabular} & 64 (2.8\%)                          \\ \hline
8 & Some votes may still need to be counted                                                                                                & 26 (1.1\%)                          \\ \hline
9 & Get the facts about COVID-19                                                                                                           & 26 (1.1\%)                          \\ \hline
10 & Esta reivindicação de fraude é contestada                                                                                              & 11 (0.5\%)                          \\ \hline
11 & Saiba por que urnas eletrônicas são seguras                                                                                            & 11 (0.5\%)                          \\ \hline
12 & Sources called this election differently                                                                                               & 3 (0.1\%)                           \\ \hline
13 & Get the facts about mail-in ballots                                                                                                    & 2 (0.1\%)                           \\ \hline
\end{tabular}%
}
\caption{Warning labels in our dataset.}
\label{tab:all-labels}
\end{table}

\subsection{Warning Labels}

We start by looking into the different types of warning labels that exist in our dataset.
To do this, we focus on tweets that include warning labels (see the first row in Table~\ref{tab:dataset}), specifically, 2,244 tweets posted by 853 users. 

Table~\ref{tab:all-labels} shows all warning labels in our dataset and their respective frequency and percentage over all the tweets.
Overall, we find 13 different warning labels with the majority of them being related to the 2020 US elections. 
For instance, the most popular warning label in our dataset is ``This claim about election fraud is disputed'' with 58\% of all tweets.
Other 2020 US election warning labels are related to the security of the elections like ``Learn about US 2020 election security efforts'' (12\%) and ``Learn how voting by mail is safe and secure'' (5.8\%), as well as related to the outcome of the elections like ``Multiple sources called this election differently'' (4.2\%) and ``Election officials have certified Joe Biden as the winner of the U.S. Presidential election'' (2.8\%).
Interestingly, we also find warning labels referring to the 2020 US elections written in other languages (i.e., Portuguese).
We find 0.49\% tweets including ``Esta reivindicação de fraude é contestada'' (translates to ``This fraud claim is disputed'') and ``Saiba por que urnas eletrônicas são seguras'' (translates to ``Find out why electronic voting machines are safe'').
We manually examine the tweets that had warning labels in Portuguese, finding that all tweets were posted in the Portuguese language by 15 different users (the language of the warning label is independent of the user's language preference on Twitter).
Apart from politics-related warning labels, we find a general-purpose warning label that informs users about manipulated media (images or videos) with 8.7\% of all tweets in our dataset.
Finally, we find a COVID-19 specific warning label: ``Get the facts about COVID-19'' (1.15\%) that aims to inform users about health-related issues and, in particular, the COVID-19 pandemic.
The small percentage of tweets with COVID-19 related warning labels is likely because our dataset is mainly centered around the time period of the 2020 US elections.

\begin{figure*}[t!]
\centering
 \includegraphics[width=0.75\textwidth]{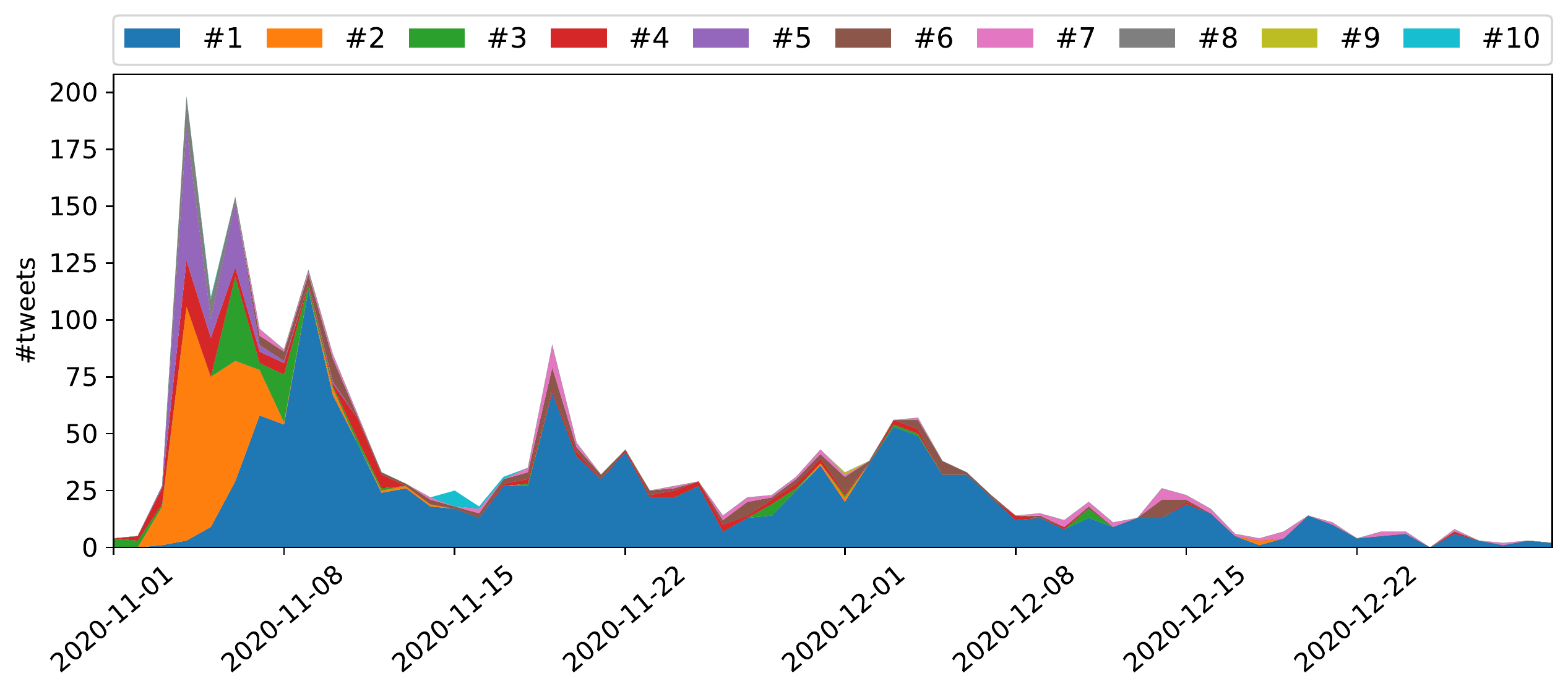}
\caption{Number of tweets that include a warning label for each day between November 1, 2020 and December 30, 2020 (for the top 10 warning labels). See Table~\ref{tab:all-labels} for mapping the warning label text to the numbers in the legend.} 
\label{fig:warnings_over_time}
\end{figure*}

Next, we analyze how these warning labels appear over time on Twitter.
Fig.~\ref{fig:warnings_over_time} shows how the top 10 most popular warning labels in our dataset appear over time on Twitter.
We focus on the period between November 2020 and December 2020 for readability purposes and because most of the tweets in our dataset are shared during this period (see Section~\ref{sec:dataset}).
We find two different temporal patterns.
First, we find warning labels that are short-lived as most of their appearances on tweets happen within a short period of time.
Concretely, both ``Learn about US 2020 election security efforts'' (\#2) and ``Official sources may not have called the race when this was Tweeted'' (\#5) are exclusively used during the first week of November 2020.
On the other hand, we find warning labels that are long-lived.
E.g., the label ``This claim about election fraud is disputed'' (\#1) is used for the entirety of the period between November and December 2020. 
Another example is ``Manipulated media'' (\#3) used during the entire period (from March 2020 to December 2020).
Overall, these results indicate that warning labels are time and context-dependent, with some of them being short-lived (few days) and some of them being long-lived (several months).

\begin{figure}[t!]
\centering
\includegraphics[width=0.85\columnwidth]{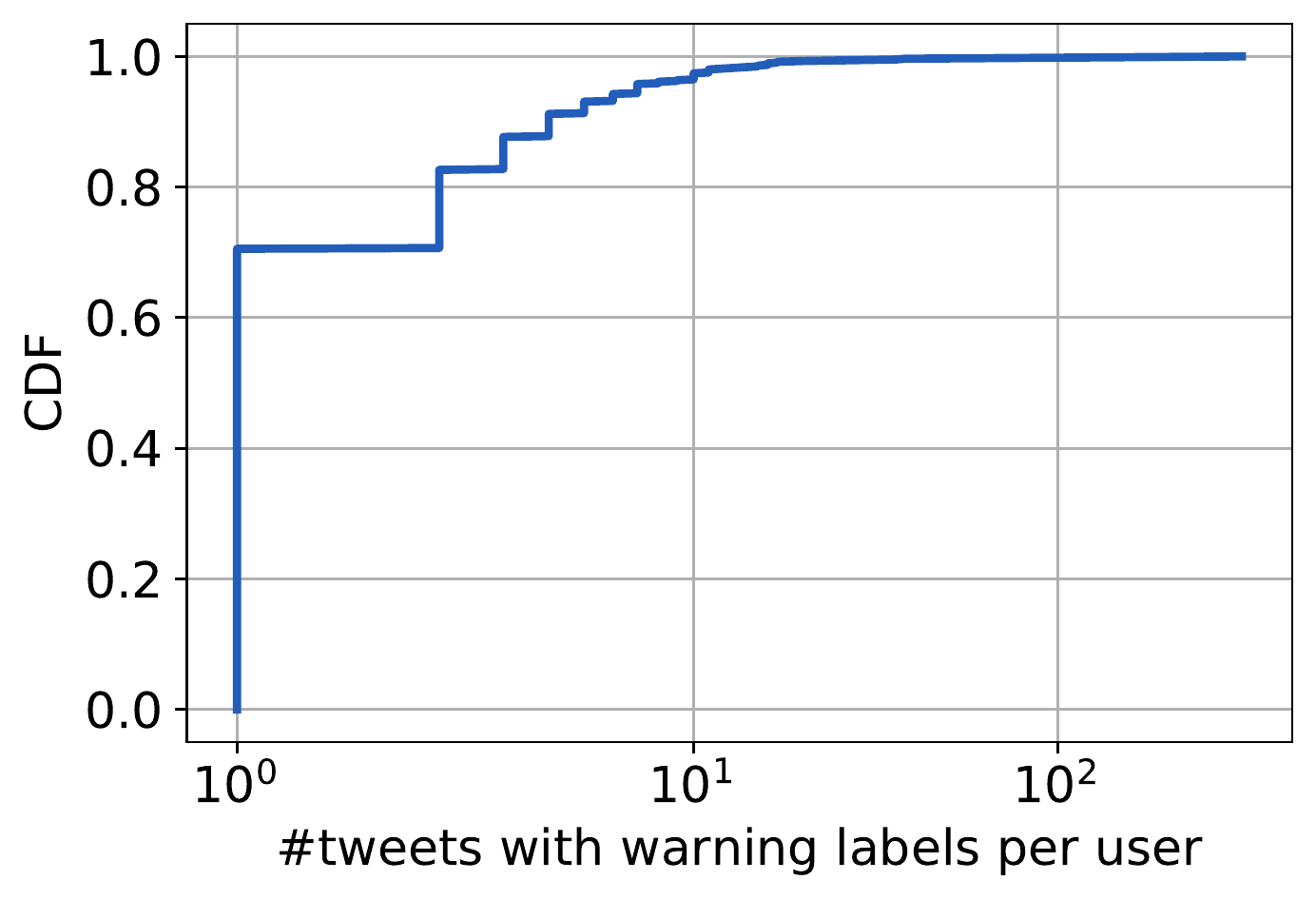}
\caption{Cumulative Distribution Function (CDF) of the number of tweets with warning labels per user.} 
\label{fig:cdf_warnings_per_user}
\end{figure}

\begin{table*}[t]
\centering
\resizebox{0.9\textwidth}{!}{%
\begin{tabular}{@{}lrrrrrr@{}}
\toprule
\textbf{}           & \multicolumn{2}{c}{\textbf{All}}                                   & \multicolumn{2}{c}{\textbf{Tweets with warning labels}}                                     & \multicolumn{2}{c}{\textbf{Quoted tweets}}                               \\ \midrule
                    & \multicolumn{1}{c}{\textbf{Tweets}} & \multicolumn{1}{c}{\textbf{Users}}  & \multicolumn{1}{c}{\textbf{Tweets}} & \multicolumn{1}{c}{\textbf{Users}}         & \multicolumn{1}{c}{\textbf{Tweets}} & \multicolumn{1}{c}{\textbf{Users}} \\ \cmidrule(l){2-7} 
\textbf{Republicans} & 4,232 (22.5\%)                      & \multicolumn{1}{r|}{1,095 (13.4\%)} & \textbf{1,634 (72.8\%)}             & \multicolumn{1}{r|}{\textbf{404 (47.3\%)}} & 2,775 (16.5\%)                      & 917 (11.9\%)                       \\
\textbf{Neutral}    & 4,610 (24.5\%)                      & \multicolumn{1}{r|}{2,604 (31.9\%)} & 294 (13.1\%)                        & \multicolumn{1}{r|}{184 (21.5\%)}          & 4,347 (25.9\%)                      & 2,483 (32.3\%)                     \\
\textbf{Democrats}   & 9,747 (51.9\%)                       & \multicolumn{1}{r|}{4,324 (53.1\%)} & 262 (11.6\%)                        & \multicolumn{1}{r|}{218 (25.5\%)}          & \textbf{9,491 (56.6\%)}             & \textbf{4,198 (54.7\%)}            \\
\textbf{N/A}        & 176 (0.9\%)                         & \multicolumn{1}{r|}{119 (1.4\%)}    & 54 (2.4\%)                          & \multicolumn{1}{r|}{47 (5.5\%)}            & 127 (0.7\%)                         & 75 (0.9\%)                         \\ \bottomrule
\end{tabular}%
}
\caption{Inferred political leaning of users who shared tweets with warning labels or quoted a tweet that had a warning label.}
\label{tab:leanings}
\end{table*}

\subsection{User Analysis}

Here, we look into the users who share tweets with warning labels.
Recall that our data collection involves 168K users, and only 853 of them share tweets that have warning labels, hence indicating that only a small percentage (0.5\%) of Twitter verified users have warning labels attached to their content.

As per Fig.~\ref{fig:cdf_warnings_per_user}, out of the 853 users, 70\% had only one tweet with a warning label, while only 3.6\% of these users had at least 10 tweets with warning labels.
Overall, only a small percentage of users have warning labels on multiple of their tweets.

\begin{table}[t!]
\centering
\resizebox{0.95\columnwidth}{!}{%
\begin{tabular}{@{}lllrr@{}}
\toprule
\multicolumn{1}{c}{\textbf{User}} & \multicolumn{1}{c}{\textbf{\begin{tabular}[c]{@{}c@{}} Political\\ leaning\end{tabular}}} & \multicolumn{1}{c}{\textbf{\begin{tabular}[c]{@{}c@{}}Account \\status\end{tabular}}} & \multicolumn{1}{c}{\textbf{Tweets}}  \\ \midrule
realDonaldTrump                                                    & Republican                                                                                        & Suspended                                                                                                     & 321 (14.3\%)                                                                                                           \\
TeamTrump                                                          & Republican                                                                                        & Suspended                                                                                                     & 105 (4.6\%)                                                                                                            \\
gatewaypundit                                                      & Republican                                                                                        & Active                                                                                                        & 71 (3.1\%)                                                                                                             \\
va\_shiva                                                          & Neutral                                                                                           & Active                                                                                                        & 38 (1.6\%)                                                                                                            \\
JudicialWatch                                                      & Republican                                                                                        & Active                                                                                                        & 36 (1.6\%)                                                                                                            \\
MichaelCoudrey                                                     & Republican                                                                                        & Suspended                                                                                                     & 27 (1.2\%)                                                                                                          \\
TomFitton                                                          & Republican                                                                                        & Active                                                                                                        & 20 (0.8\%)                                                                                                            \\
RudyGiuliani                                                       & Republican                                                                                        & Active                                                                                                        & 17 (0.7\%)                                                                                                              \\
JamesOKeefeIII \textbf{(U)}                       & Republican                                                                                        & Active                                                                                                        & 17 (0.7\%)                                                                                                             \\
EmeraldRobinson                                                    & Republican                                                                                        & Active                                                                                                        & 16 (0.7\%)                                                                                                             \\
RealJamesWoods                                                     & Republican                                                                                        & Active                                                                                                        & 16 (0.7\%)                                                                                                           \\
LLinWood \textbf{(U)}                             & Republican                                                                                        & Suspended                                                                                                     & 16 (0.7\%)                                                                                                           \\
realLizUSA                                                         & Republican                                                                                        & Active                                                                                                        & 15 (0.6\%)                                                                                                            \\
LouDobbs                                                           & Republican                                                                                        & Active                                                                                                        & 15 (0.6\%)                                                                                                             \\
KMCRadio                                                           & Republican                                                                                        & Suspended                                                                                                     & 14 (0.6\%)                                                                                                              \\
michellemalkin                                                     & Republican                                                                                        & Active                                                                                                        & 13 (0.5\%)                                                                                                             \\
CodeMonkeyZ \textbf{(U)}                          & Republican                                                                                        & Suspended                                                                                                     & 12 (0.5\%)                                                                                                              \\
charliekirk11                                                      & Neutral                                                                                           & Active                                                                                                        & 11 (0.4\%)                                                                                                           \\
TrumpWarRoom                                                       & Republican                                                                                        & Active                                                                                                        & 11 (0.4\%)                                                                                                          \\
chuckwoolery                                                       & Republican                                                                                        & Active                                                                                                        & 11 (0.4\%)                                                                                                           \\ \bottomrule
\end{tabular}%
}
\caption{Top 20 users who had the most warning labels on their tweets. (U) refers to unverified users who exist in our dataset because verified users retweeted or quoted tweets of them that had warning labels. We also report the account status of each user as of January 9, 2021. Note that we do not anonymize any username because all users are public figures (including the three unverified users).}
\label{tab:top-20-users}
\end{table}

\descr{Users' political leaning.} As we described above, our dataset has a strong political nature, and the majority of the warning labels refer to claims about the 2020 US elections (e.g., claims about election fraud, see Table~\ref{tab:all-labels}).
Motivated by this, we augment our dataset with information about the political leaning of each user that shared tweets with warning labels.
To infer users' political leaning, we use the methodology presented by~\cite{kulshrestha2017quantifying,kulshrestha2019search} and, in particular, the Political Bias Inference API that is publicly available by~\cite{inference_api}. 
The API generates a vector with the topical interests of each user and their frequency. 
To do this, the API collects all the user's friends (i.e., people that the user follows), generates all the topics inferred for each friend using the methodology in~\cite{ghosh2012cognos,sharma2012inferring}, and calculates a vector with all the topics and their frequencies.
Finally, by comparing the topical vectors to a ground truth dataset of Republican and Democrat Twitter users, the API infers whether a Twitter user has a Republican, Democrat, or Neutral political leaning.

In this work, we use the Political Bias Inference API, between January 3 and January 10, 2021, to infer the political leaning of the 8,142 Twitter users in our dataset.
We quantify the performance of the Political Bias Inference API in our dataset by manually classifying 200 users from our dataset, finding 80.5\% accuracy (see Appendix~\ref{sec:appendix}).
Table~\ref{tab:leanings} reports the number of tweets and users per inferred political leaning for the entire dataset and broken down into tweets with warning labels and quoted tweets.
We observe that for the entire dataset, 51\% of the users are Democrats, 13.4\% are Republicans, almost 32\% are inferred as neutral, while for the rest 1.4\% we were unable to infer their political leaning. 
This is because some users were either suspended or made their accounts private by the time we were collecting their friend list, hence the Political Bias Inference API was unable to make an inference.

Interestingly, when looking at the tweets with warning labels in Table~\ref{tab:leanings}, we find that most of the tweets with warning labels are shared by Republicans (72\% of all tweets vs. 11\% for Democrats).
This likely indicates that due to the context and developments related to the 2020 US elections, Republicans tend to share more questionable content that is more likely to receive warning labels by Twitter.
For the quoted tweets, we observe that Democrats tend to comment on tweets with warning labels more often than Republicans (56\% vs. 16.5\% for Republicans).

\begin{table}[t!]
\centering
\resizebox{0.95\columnwidth}{!}{%
\begin{tabular}{@{}llll@{}}
\toprule
\multicolumn{1}{c}{\textbf{User}} & \multicolumn{1}{c}{\textbf{\begin{tabular}[c]{@{}c@{}}Political\\ leaning\end{tabular}}} & \multicolumn{1}{c}{\textbf{\begin{tabular}[c]{@{}c@{}}Account\\  status\end{tabular}}} & \multicolumn{1}{c}{\textbf{Tweets}} \\ \midrule
realDonaldTrump                          & Republican                                                                               & Suspended                                                                              & 78 (0.4\%)                          \\
AndrewFeinberg                           & Democrat                                                                                 & Active                                                                                 & 52 (0.3\%)                          \\
svdate                                   & Neutral                                                                                  & Active                                                                                 & 50 (0.3\%)                          \\
NumbersMuncher                           & Republican                                                                               & Active                                                                                 & 49 (0.3\%)                          \\
atrupar                                  & Democrat                                                                                 & Active                                                                                 & 44 (0.2\%)                          \\
GlennKesslerWP                           & Democrat                                                                                 & Active                                                                                 & 42 (0.2\%)                          \\
T\_S\_P\_O\_O\_K\_Y                      & Republican                                                                               & Active                                                                                 & 42 (0.2\%)                          \\
BrianKarem                               & Democrat                                                                                 & Active                                                                                 & 39 (0.2\%)                          \\
Patterico                                & Republican                                                                               & Active                                                                                 & 38 (0.2\%)                          \\
TalbertSwan                              & Neutral                                                                                  & Active                                                                                 & 37 (0.2\%)                          \\
BarnettforAZ                             & Republican                                                                               & Active                                                                                 & 33 (0.2\%)                          \\
TomFitton                                & Republican                                                                               & Active                                                                                 & 32 (0.2\%)                          \\
Justin\_Stangel                          & Democrat                                                                                 & Active                                                                                 & 31 (0.2\%)                          \\
JLMarchese111                            & N/A                                                                                      & Suspended                                                                              & 31 (0.2\%)                          \\
HalSparks                                & Democrat                                                                                 & Active                                                                                 & 30 (0.2\%)                          \\
RhondaFurin                              & Republican                                                                               & Active                                                                                 & 29 (0.2\%)                          \\
captainjanks                             & Democrat                                                                                 & Active                                                                                 & 28 (0.2\%)                          \\
rogerkimball                             & Republican                                                                               & Active                                                                                 & 27 (0.2\%)                          \\
amhfarraj                                & Democrat                                                                                 & Active                                                                                 & 26 (0.2\%)                          \\
michellemalkin                           & Republican                                                                               & Active                                                                                 & 25 (0.1\%)                          \\ \bottomrule
\end{tabular}%
}
\caption{Top 20 users who quoted tweets that had warning labels. We also report the account status of each user as of January 9, 2021.}
\label{tab:top-20-users-quoted}
\end{table}

\descr{Top users.} But who are the users who are the most ``prolific'' with regards to tweets that include warning labels or in the quoted tweets?
Table~\ref{tab:top-20-users} and Table~\ref{tab:top-20-users-quoted} show the top 20 users in our dataset based on the number of tweets with warning labels and the quoted tweets, respectively.
For each user, we report the inferred political leaning and whether the account was active or suspended on January 9, 2021.
We make several observations.
First, in both cases, the most prolific user is President Trump, with 14.3\% of all tweets with warning labels and 0.4\% of all quoted tweets.
The account of President Trump was permanently suspended by Twitter on January 8, 2021, due to the risk of further incitement of violence~\cite{twitter_trump_suspension}, after his supporters attacked the US capitol~\cite{capitol_attack}.
Second, we observe that most of the top 20 users who shared tweets with warning labels are inferred as Republicans (see Table~\ref{tab:top-20-users}).
This is not the case for the quoted dataset (see Table~\ref{tab:top-20-users-quoted}), as 8 out of the top 20 users with quoted tweets are inferred as Democrats.
Third, even though our study does not focus on unverified users, we observe the existence of three unverified accounts among the top 20 users who shared tweets with warning labels.\footnote{This is because we collect the tweets that verified accounts retweeted or quoted by unverified accounts (see Section~\ref{sec:dataset}).}
This indicates that Twitter's moderation mechanism is not only limited to verified users.
Finally, we note that 6 out of the top 20 users with tweets that had warning labels were suspended by Twitter (as of January 9, 2021).
This highlights that the continuous dissemination of questionable content that leads to the addition of warning labels is likely to result in hard moderation interventions (i.e., user suspensions).

\begin{figure*}[t!]
\centering
\subfigure[]{\includegraphics[width=0.4\textwidth]{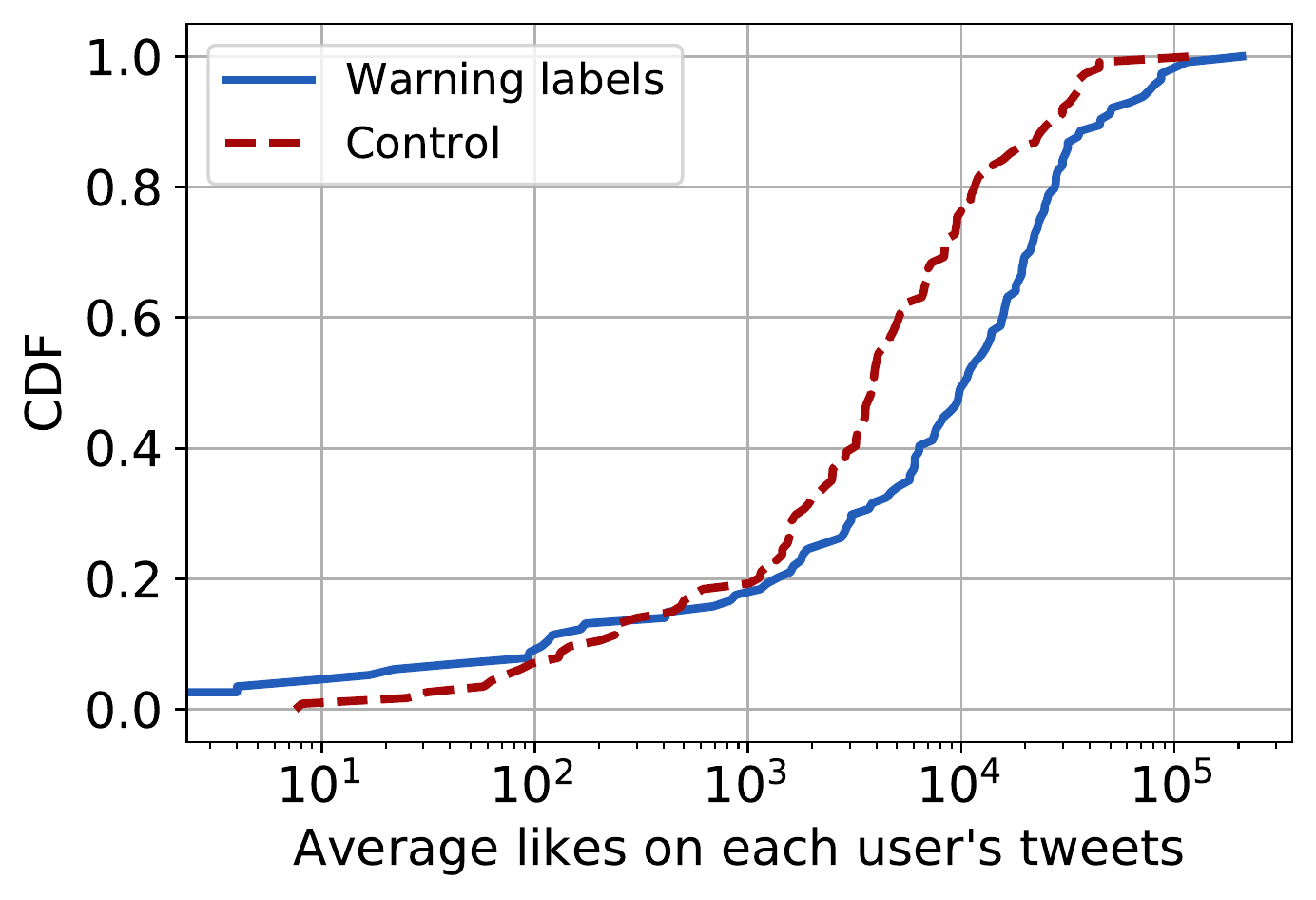}\label{fig:cdf_favorite}}
\subfigure[]{\includegraphics[width=0.4\textwidth]{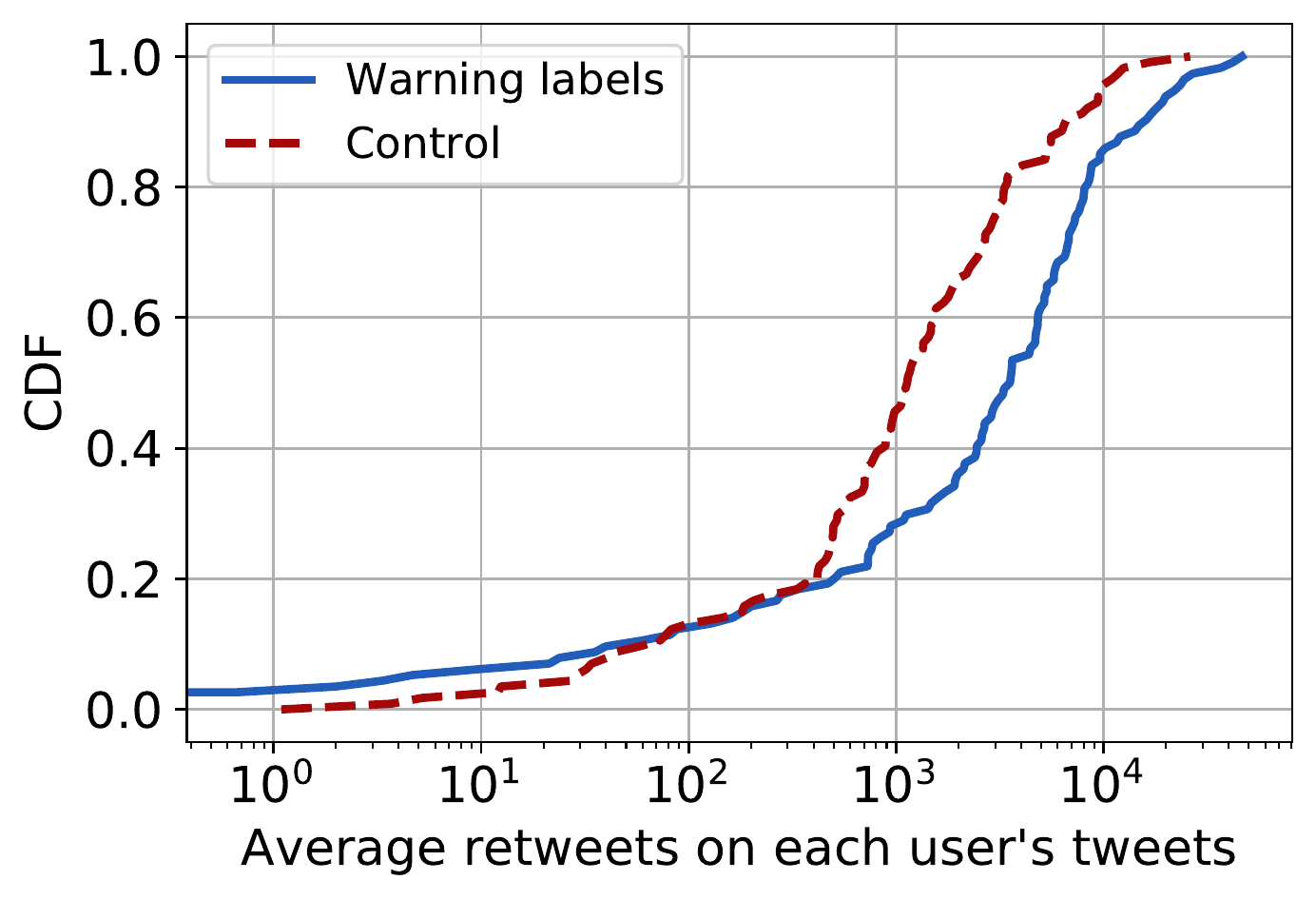}\label{fig:cdf_retweets}}
\subfigure[]{\includegraphics[width=0.4\textwidth]{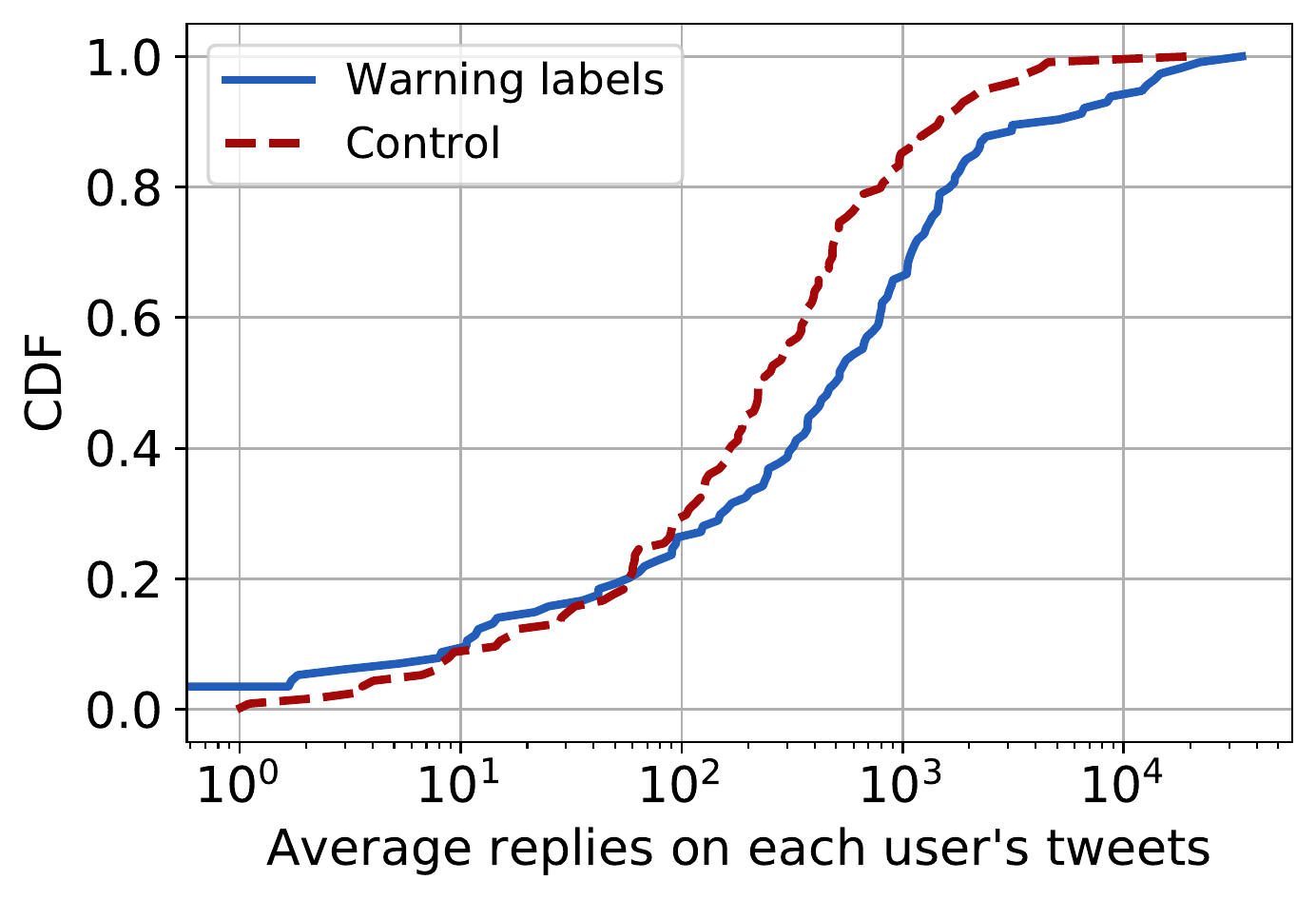}\label{fig:cdf_replies}}
\subfigure[]{\includegraphics[width=0.4\textwidth]{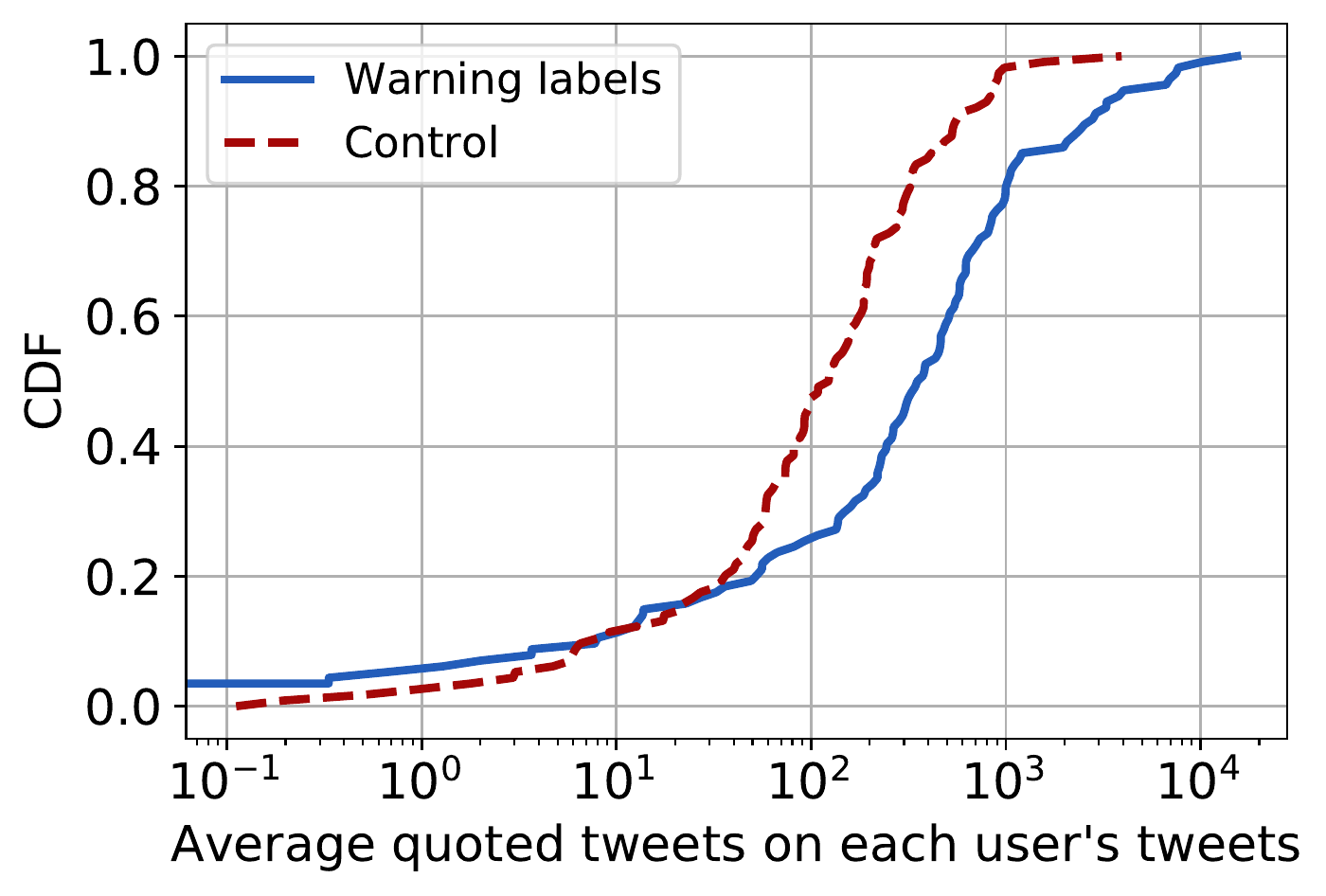}\label{fig:cdf_quotes}}
\caption{Mean engagement metric for each user for tweets with warning labels and for tweets without warning labels.}
\label{fig:engagement_analysis}
\end{figure*}

\descr{Take-aways.} The main take-away points from our analysis on warning labels and Twitter users are:

\begin{compactenum}
    \item Most of the warning labels on Twitter, between November 2020 and December 2020, were related to the 2020 US elections. Also, we find different temporal patterns in the use of warning labels, with a few of them being short-lived (less than a week) and some of them being long-lived (across several months).
    \item We find warning labels used to inform users about manipulated multimedia, while some warning labels are in languages other than English (i.e., Portuguese). This highlights the efforts put in soft moderation interventions and some of the challenges (e.g., tracking claims across multiple information formats or languages).
    \item The majority of tweets with warning labels (72\%) are shared by Republicans, while Democrats are more likely to comment on tweets with warning labels using Twitter's quoting functionality (56\% of the tweets compared to 16\% for Republicans). These results likely indicate that Republicans, during the 2020 US elections, shared more questionable content that was eventually flagged by Twitter moderators.
    \item The continuous dissemination of potentially harmful information that is annotated with warning labels can lead to hard moderation interventions like permanent user suspensions. We find that 6 out of the 20 top users, in terms of sharing tweets with warning labels, were permanently suspended by Twitter as of January 9, 2021.
\end{compactenum}

\section{RQ2: Engagement Analysis}
The goal of warning labels is to provide adequate information on tweets that include questionable content and might be harmful to users or society.
Thus, we expect that users who see content that is annotated with warning labels are likely to cause them to be less willing to engage with or reshare such content~\cite{mena2020cleaning}. 
In this section, we aim to quantify the differences in the engagement between tweets that include warning labels and tweets that do not.
Our empirical analysis can quantify how effective are the warning labels on Twitter through the lens of user engagement.

\begin{figure}[t!]
\centering
\includegraphics[width=0.85\columnwidth]{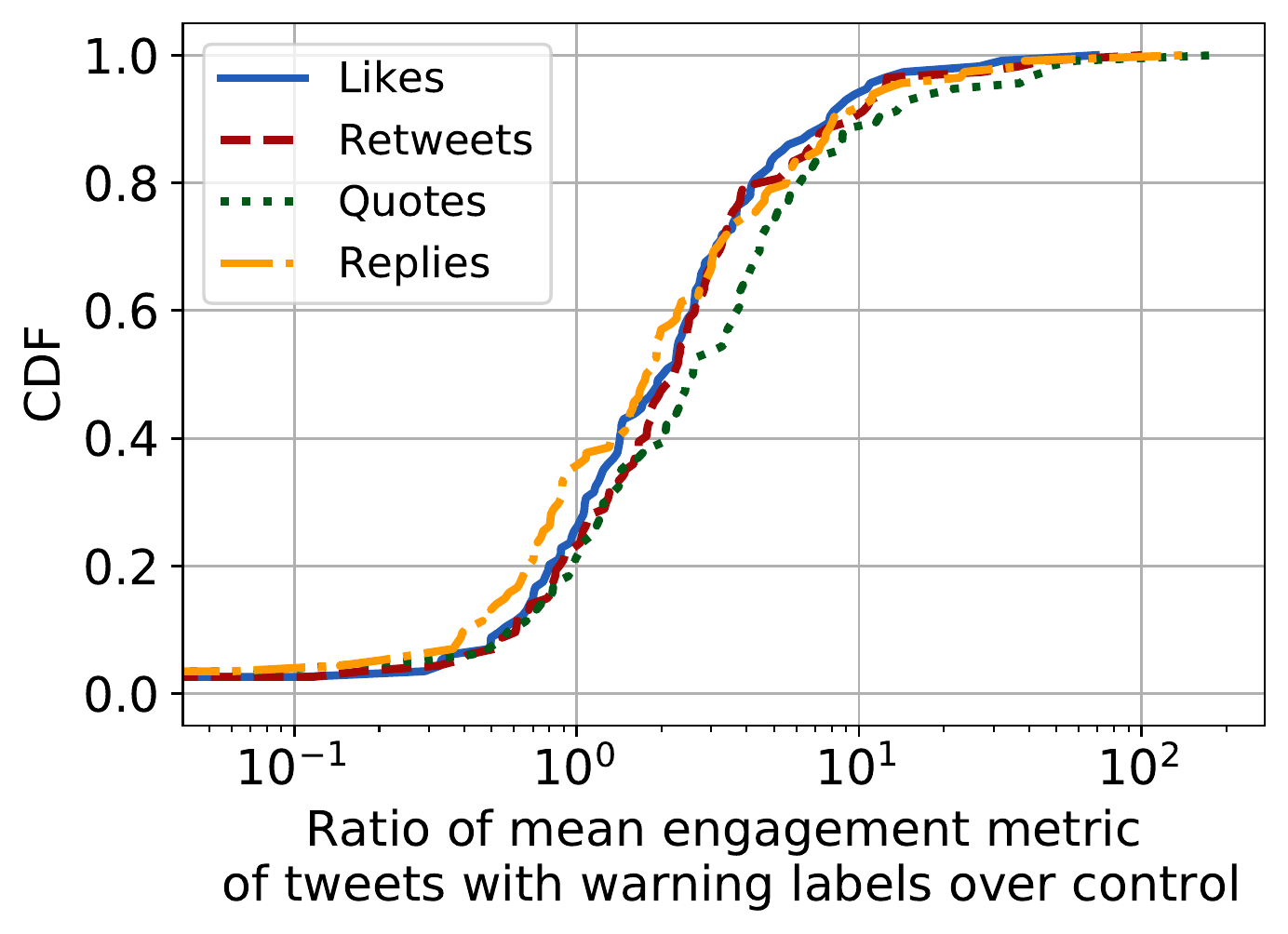}\label{fig:mean_diffs}
\caption{CDF of the ratio of the mean engagement metric for tweets with warning labels over tweets without warning labels.}
\label{fig:mean_diffs_per_user}
\end{figure}

\begin{figure*}[t!]
\centering
\subfigure[]{\includegraphics[width=0.4\textwidth]{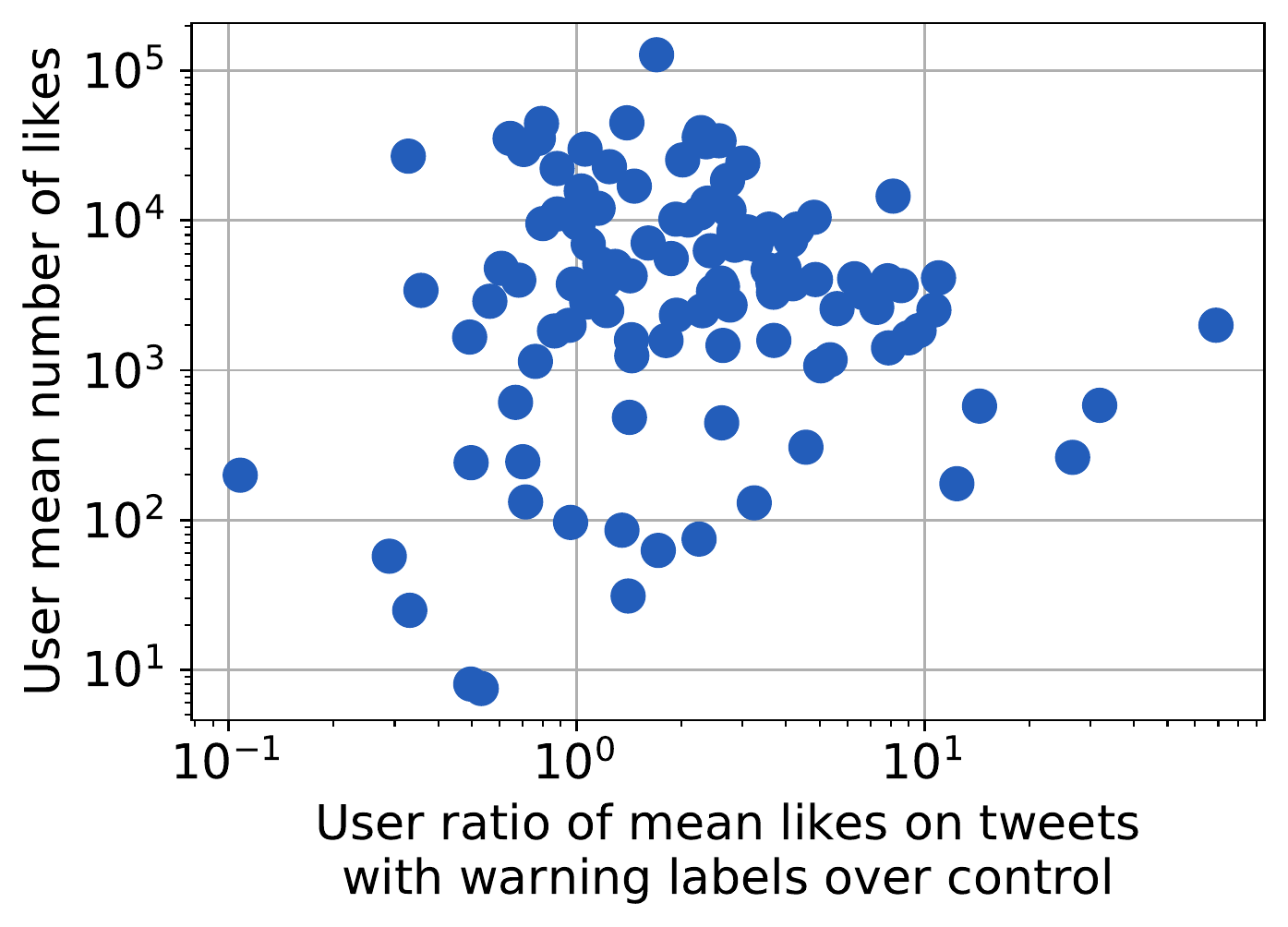}\label{fig:scatter_favorite}}
\subfigure[]{\includegraphics[width=0.4\textwidth]{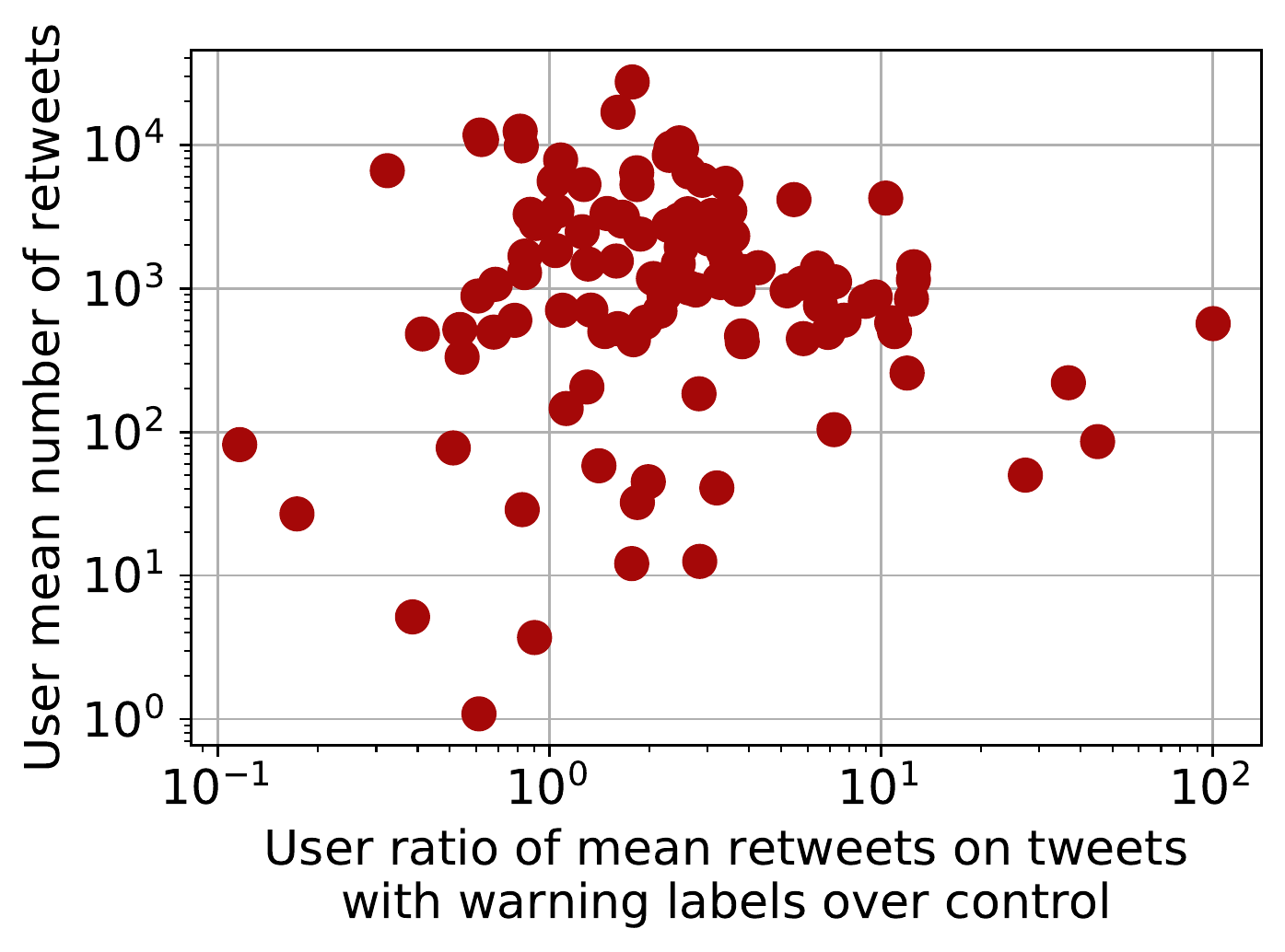}\label{fig:scatter_retweets}}
\subfigure[]{\includegraphics[width=0.4\textwidth]{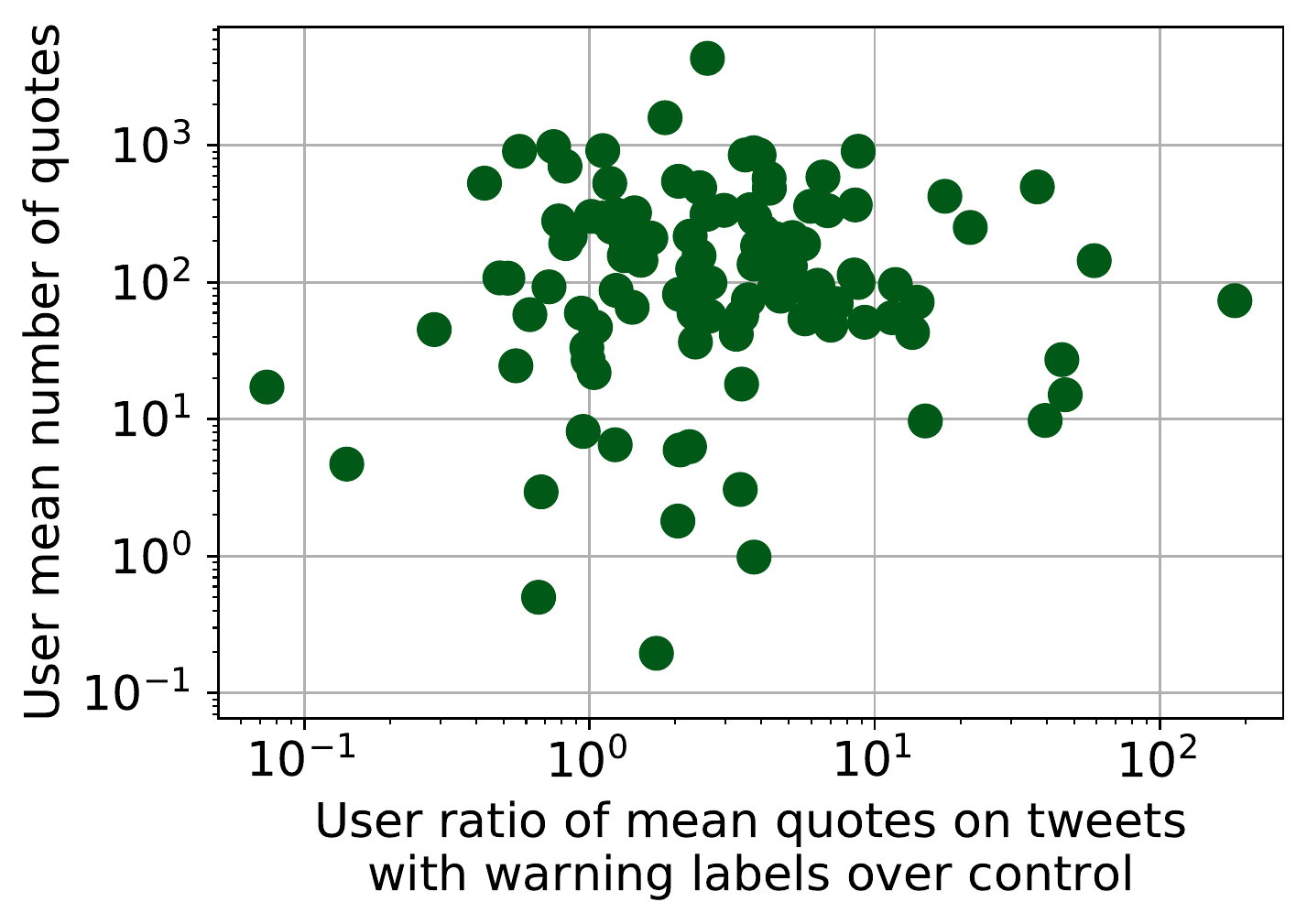}\label{fig:scatter_replies}}
\subfigure[]{\includegraphics[width=0.4\textwidth]{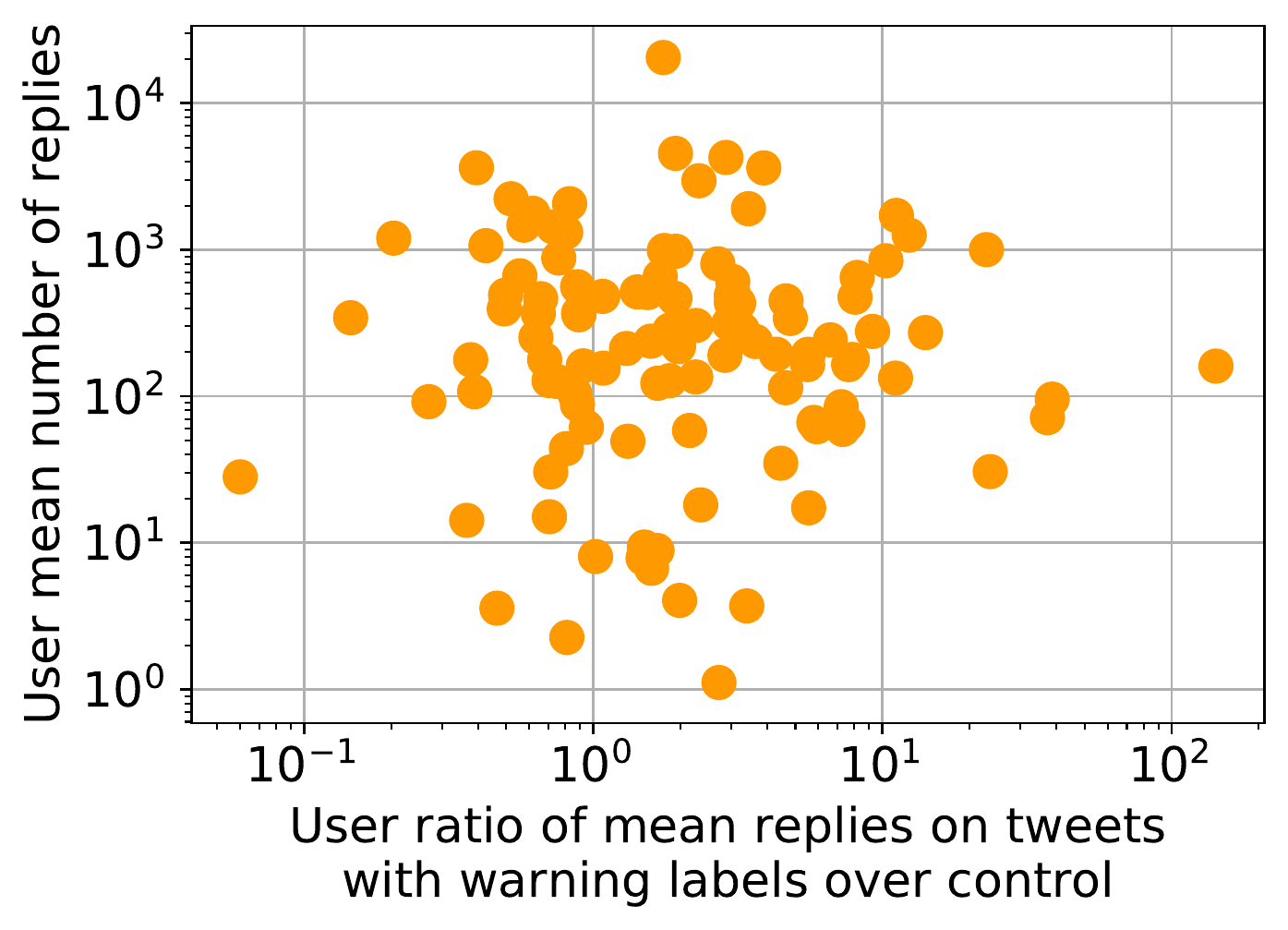}\label{fig:scatter_quotes}}
\caption{Number of user followers and the ratio of the mean engagement metric for tweets with warning labels over the control dataset.}
\label{fig:popularity_change_in_engagement}
\end{figure*}

For each user in our dataset, we extract two sets of tweets:
1)~tweets that have warning labels;
2)~a control dataset of tweets that do not have warning labels.
Note that we limit our analysis to the 115 users with at least three tweets with warning labels to make sure that our analysis is not influenced by one or two tweets.
Then for each engagement metric in our dataset, we calculate the mean number that each group of tweets (warning label tweets and control) had for each user. 
Our analysis takes into account four engagement metrics:
1)~\emph{Likes} (how many times the tweet was liked by other users);
2)~\emph{Retweets} (how many times the tweet was retweeted by other users);
3)~\emph{Quotes} (number of other tweets that retweeted the tweet with a comment); 
and 4)~\emph{Replies} (number of replies that the tweet received).

Fig.~\ref{fig:engagement_analysis} shows the Cumulative Distribution Function (CDF) of the average number of likes/retweets/quotes/replies of tweets with and without warning labels per user.
For each engagement metric, we perform two-sample Kolmogorov-Smirnov statistical significance tests, finding that in all cases, the engagement of tweets with warning labels is significantly different compared to tweets without warning labels ($p<0.01$) 
We observe that, for all four engagement metrics, users receive increased engagement on tweets that have warning labels.

For likes (see Fig.~\ref{fig:cdf_favorite}), we find a median value of 10,303.9 average likes per user for tweets with warning labels, whereas, for the control dataset, we find a median value of 3,834.3 (2.6x less than warning labels). 
For retweets (see Fig.~\ref{fig:cdf_retweets}), we find a median value of 3,533 average retweets per user for tweets with warning labels, while for the control dataset, the median value is only 1,129.2 (3.1x decrease compared to the warning labels).
For replies (see Fig.~\ref{fig:cdf_replies}), we find a median value of 235.7 replies for the control dataset, while for warning labels, the median value increases to 494 (2.1x increase over the control dataset).
For quotes (see Fig.~\ref{fig:cdf_quotes}), we find a median value of 350.6 average quotes per user for the warning labels datasets, whereas, for the control dataset, we find a median value of 122.9 quotes (2.8x decrease compared to warning labels).
Also, from Fig.~\ref{fig:engagement_analysis}, we can observe a small proportion of users who have less engagement on the warning labels dataset.
To quantify the proportion of users who have more engagement on control tweets over the tweets that had warning labels, we plot the ratio of the mean number of each engagement metric on tweets with warning labels over the control dataset (see Fig.~\ref{fig:mean_diffs_per_user}).
When this ratio is below 1, user's control dataset had more engagement compared to the user's warning labels dataset. 
We find that 26\%, 23\%, 21\%, 35\% of the users had more engagement on their control tweets over the ones with warning labels for likes, retweets, quotes, and replies, respectively.

From our analysis thus far, it is unclear which set of users have increased vs. decreased engagement on tweets with warning labels over the control dataset.
To assess whether there is a correlation between the overall engagement that user receives and whether a user will receive increased or decreased engagement on tweets with warning labels, we plot the overall engagement (i.e., mean engagement metric for all the user's tweets) and the ratio of engagement on warning labels over the control dataset (see Fig.~\ref{fig:popularity_change_in_engagement}).
We observe that for all engagement metrics, most of the users that have on average high engagement on their content (i.e., over 1K likes, over 100 retweets, over 100 quotes, and over 100 replies) also receive an increased engagement on tweets with warning labels over the control (note that the ratio for these users is in most of the times between 1 and 10).

\descr{Take-aways.} The key take-away points from our engagement analysis are:
\begin{compactenum}
    \item Tweets with warning labels had more engagement compared to tweets without warning labels. This is likely because Twitter prioritizes content that receives high engagement within their soft moderation system~\cite{twitter_softmod}. 
    \item We find that 65\%-79\% (depending on engagement metric) of the users receive increased engagement on their tweets with warning labels compared to tweets without warning labels.
    \item By looking at the users that have increased vs. decreased engagement on tweets with warning labels compared to the control dataset, we find that most users who have high engagement in general have also increased engagement on tweets with warning labels.
\end{compactenum}

\section{RQ3: User Interaction with Tweets and Warning Labels}

In this section, we study how users interact with tweets that have warning labels.
To do this, we use Twitter's quote functionality, where users can retweet a tweet with a comment.
Specifically, we qualitatively analyze three sets of tweets;
1)~the 50 tweets that quote other tweets and Twitter includes warning labels on \textbf{both} tweets;
2)~122 tweets (out of the 169) that quote other tweets and Twitter includes a warning label \textbf{only} on the top tweet (i.e., user's comment). The 47 other tweets had a quoted tweet that was deleted when we tried to assess them qualitatively;
3)~150 randomly selected tweets that quote another tweet that includes a warning label.
We qualitatively analyze all three sets of tweets to understand how users interact with people that share content that is annotated with warning labels, how users interact with questionable content (e.g., false claims), and how users discuss or perceive the existence of warning labels on Twitter.
Specifically, we employ qualitative thematic analysis~\cite{braun2006using} to identify a set of codes from our dataset, code the tweets in our samples, and then group the tweets into high-level themes.
Below, we present the main themes for each sample.

\subsection{Quoted tweets where both tweets include warning labels.}

\begin{figure}[t!]
\centering
\includegraphics[width=0.85\columnwidth]{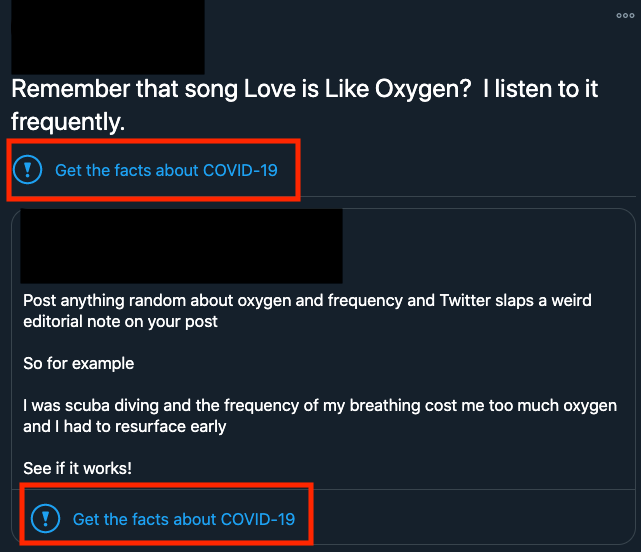}
\caption{Example of an incorrect addition of warning labels on Twitter.}
\label{fig:mislabel_example}
\end{figure}

Intuitively, when both the quoted tweet and the comment tweet above include warning labels (e.g., Fig.~\ref{fig:mislabel_example}), one expects that both tweets include information that is questionable or potentially harmful. 
Here, we qualitatively analyze the tweets in our dataset to verify if this is the case and what are other cases where both the quoted and the comment tweet include warning labels.

\descr{Reinforcing false claims.} The majority of the comments above the quoted tweets aim to retweet and reinforce the false claim included in the quoted tweet (86\%, 43 out of the 50). 
Two of them achieve this using a single word (``this'' or ``true''), two of them use videos, five of them achieve it by tweeting a single hashtag (\#stopthesteal and \#ExposeDominion that both refer to election fraud claims), while the rest of the comment tweets use text to reinforce the claim.
The fact that some of the comments comprise only a single word shows that adding warning labels to tweets requires considering the context and other quoted tweets and not focusing on the tweet in isolation.
Also, in 2 out of the 43 comments that reinforce the claim of the quoted tweet, the users share their anti-censorship opinions or disputing the fact that the content should be labeled (i.e., \emph{``Say NO to Big Tech censorship!''} and \emph{``Twitter labeled this tweet as disputed.... What exactly is Twitter disputing here?''}). 
These results further compound the findings from~\cite{saltz2020encounters}.

\descr{Testing warning labels.} We find one tweet where the user commented with the same content as the quoted tweet, likely to verify if his comment will eventually get a warning label.

\descr{Incorrect warning labels.} We find one specific case where the warning labels were seemingly incorrectly put (see Fig.~\ref{fig:mislabel_example}). Both the comment and the quoted tweet had the warning label ``Get the facts about COVID-19'' and both were including the terms oxygen and frequency/frequently. This likely indicates that Twitter employs automated means to attach warning labels and in some cases, warning labels are incorrectly added to some content.

\begin{figure}[t!]
\centering
\includegraphics[width=0.85\columnwidth]{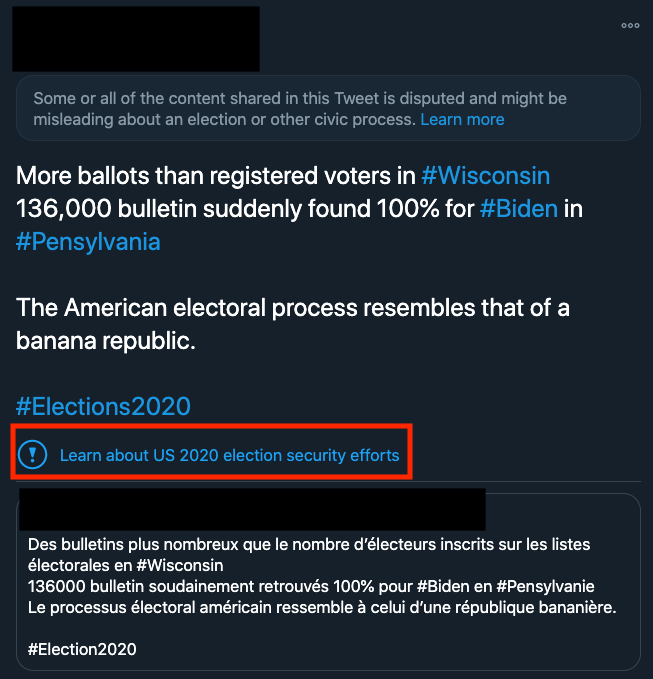}
\caption{A quoted tweet that is not flagged likely because is in French.}
\label{fig:language_example}
\end{figure}

\subsection{Warning Labels on the Comment Above the Quoted Tweet}

Next, we investigate cases where users quote a tweet with no warning label, and subsequently, their comment tweet receives a warning label (e.g., Fig.~\ref{fig:language_example}).

\descr{Making false claims.}
We find 45 tweets (36\%) that comment on real-world events, news, or facts about the election, and make false claims about the election (e.g., claims about election fraud).

\descr{Reinforcing questionable content.}
In 18 tweets (14\%), the comment above reinforces questionable content included in the quoted tweet and makes a claim even more questionable or harmful, hence getting flagged by Twitter.

\descr{Inconsistencies on warning labels.} We find several cases where there are inconsistencies with the inclusion of warning labels.
Specifically, we find 28 cases (23\%) where both the quoted tweet and the comment hint at election fraud during the 2020 US elections, yet only the quoted tweet includes a warning label.
In 7 of these cases, the comment makes a similar claim with the quoted tweet with the difference that it uses a video instead of text.
This highlights the challenges in flagging content on social media platforms and, in particular, flagging the same information across multiple diverse formats (i.e., text, images, videos).
Also, we find another case with inconsistencies related to the use of language. 
In this case, the quoted tweet and the comment above share the same information but in different languages (quoted tweet in French and comment above in English), yet only the English comment includes a warning label (see Fig.~\ref{fig:language_example}).

\begin{figure}[t!]
\centering
\includegraphics[width=0.9\columnwidth]{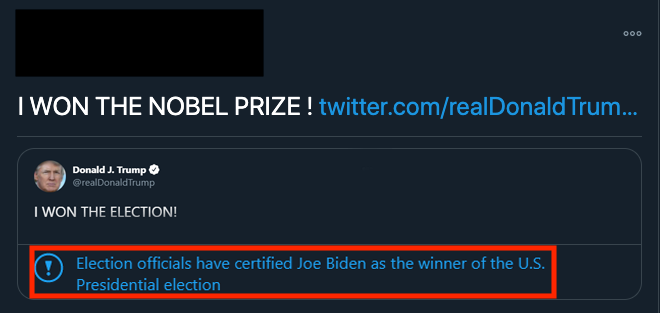}
\caption{Example of a tweet that is mocking the author or the content of the quoted tweet.}
\label{fig:trump_mock}
\end{figure}

\descr{Updates on warning labels.}
During our qualitative analysis, we observed that Twitter occasionally updates the warning labels on some tweets.
In particular, we find many instances where Twitter changed the warning label from ``Multiple sources called this election differently'' to ``Election officials have certified Joe Biden as the winner of the U.S. Presidential election''.
This highlights that Twitter continuously refines warning labels and that it is likely that they update warning labels on content to make the warning label clearer or stronger.

\subsection{Warning Labels on the Quoted Tweet}

Here, we aim to understand how users interact with content that includes warning labels by looking into tweets that quote content with warning labels (e.g., Fig.~\ref{fig:trump_mock}).
We find various behaviors ranging from mocking the author/content of the quoted tweet, debunking false claims on the quoted tweet, reinforcing the false claims, and sharing opinions on Twitter's warning labels. 
We provide more details below.

\descr{Mocking or sharing emotions about the author/content of the questionable or false claim.} We find 37 tweets that mock the content or the author of the tweet that includes a warning label. 
For instance, when President Trump tweeted the tweet in Fig.~\ref{fig:example-softmod}, several users quoted that tweet and made absurd claims about themselves like \emph{``I WON THE NOBEL PRIZE !''} (see Fig.~\ref{fig:trump_mock}) and \emph{``Let me try... I AM BEYONCE!!''}.
Other users quoted tweets with warning labels to express their emotions on the content or the author of the tweet: 4 tweets calling the quoted tweet author a liar, 4 tweets calling the author a loser, 6 tweets expressing their disgrace for the content of the tweet, and 1 tweet expressing embarrassment.

\descr{Debunking false claims.} We find 19 tweets that debunk false claims that are in quoted tweets. For instance, a user quoted a tweet shared by President Trump and wrote: \emph{``President Trump just tweeted again about claims of "secretly dumped ballots" for Biden in Michigan. This is false. 
These claims are based on screenshots of a mistaken unofficial tally on one site's election map that was caused by a typo that was corrected in about 30 minutes.''}

\descr{Reinforcing false claims.} Similarly to the tweets where both the quoted and the comment above had warning labels, we find 6 tweets that were reinforcing false claims that exist on the quoted tweets.

\descr{Sharing opinion on warning labels.} We find 6 tweets that share users' opinions on warning labels and how effective they are.
Specifically, one tweet indicates that the quoted tweet includes a warning label and two tweets question how effective the warning labels are and request stronger and more straightforward labels.
Also, we find three tweets that call for hard moderation interventions (i.e., user bans), in particular asking Jack Dorsey (Twitter's Chief Executive Officer) or Twitter Support to ban the account of President Trump due to the spread of false claims (e.g., \emph{``.@jack @Twitter make this lying stop! 
Your warnings of him lying just are not enough. 
\#BanTrump''}).
Interestingly, we find one tweet where the comment reinforces the false claim included in the quoted comment by claiming that Twitter tries to cover up the election fraud using warning labels.

\descr{Other Themes.} The rest of the tweets we qualitatively analyzed are tweets where users shared their personal or political opinion on the content of the quoted tweet or cases where users retweeted the content of the quoted tweet either by paraphrasing or translating the content to other languages.

\subsection{Take-Aways}
The main take-away points from our qualitative analysis are:
\begin{compactenum}
    \item We find various user interactions with tweets that have warning labels, such as debunking false claims, mocking users that tweeted questionable content, or reinforcing false claims despite the inclusion of warning labels.
    \item Soft moderation intervention systems are not always consistent, as we find several cases where content should have warning labels but it does not. E.g., we find cases where videos share the same information with textual tweets that include warning labels, however, the tweet with the video does not include a warning. Another example is with content across various languages. These cases show the challenges that exist on large-scale soft moderation systems.
    \item We find a case where the warning label was incorrectly added, likely due to automated means. This shows the need to devise systems that rely on human moderators that get signals from automated means (i.e., the human makes the final decision), hence decreasing the likelihood of such cases.
\end{compactenum}

\section{Discussion \& Conclusion}\label{sec:conclusion}

In this work, we performed one of the first characterizations, based on empirical data, of soft moderation interventions on Twitter.
Using a mixed-methods approach, we analyzed the warning labels, the users that share tweets that have warning labels, and the engagement that this content receives. 
Also, we investigated how users interact with such content and what are the challenges and some inconsistencies that exist on large-scale soft moderation systems.

Our user analysis showed that 72\% of the tweets with warning labels were shared by Republicans, which  likely indicates that Republicans were sharing more questionable/harmful content during the 2020 US elections.
Overall, this finding prompts the need for greater transparency by social media platforms to ease concerns related to censorship and possible moderation biases towards a specific political party~\cite{twitter_biased}.

Our engagement analysis showed that tweets with warning labels had more engagement than tweets without warning labels. 
This finding is in contrast with previous work by~\cite{mena2020cleaning} that found that users were less willing to share content with warning labels.
This likely indicates that in political discussions on Twitter, users are affected more by their own political ideology and biases rather than warning labels added by Twitter. 
Another possible explanation is that the tweets that eventually received warning labels were potentially harmful and were getting a large engagement even before the addition of the warning label.
Nevertheless, the fact that tweets with warning labels receive a substantial engagement by users, highlights the need for stricter soft moderation designs. 
Indeed, Twitter started imposing further restrictions on tweets like the ability of Twitter users to retweet, like, or reply to tweets that include content that can be extremely harmful for the platform and society at large (e.g., that may cause real-world violence)~\cite{twitter_changes}.

Finally, our qualitative analysis showed that users further debunk false claims using Twitter's quoting mechanism, they mock the user/content of the tweet with a warning label, and they reinforce false claims (despite the existence of warning labels).
Also, similarly to~\cite{saltz2020encounters}, we observed empirical evidence of users perceiving warning labels as an act of censorship by the platforms and publicly disseminating their opinion on Twitter.
Furthermore, we found some inconsistencies in content that should be flagged across multiple information formats or languages.
This highlights some of the challenges that exist when designing moderation systems that are applied to enormous and diverse platforms like Twitter.
Taken altogether, as a research community we need to further study such moderation systems to fully understand how they work and what their caveats are to increase their effectiveness, consistency, fairness, and transparency.

\descr{Limitations.} Our work has some limitations.
First, we analyzed mainly politics-related content, shared during a short period of time (two months), on a single platform (Twitter).
Thus, it is unclear whether our results hold in contexts not related to politics or to soft moderation systems that exist on other platforms like Facebook (as it has different platform affordances and design of soft moderation interventions).
Also, our engagement analysis does not account for the content of tweets, hence we do not investigate whether the increased engagement on tweets with warning labels is due to the dissemination of more controversial or sensationalistic content that is likely to attract more users.
Finally, since we do not know exactly when a soft moderation intervention happened and how the engagement changed over time, we do not analyze whether the warning labels were added because the tweets received large engagement in advance.

\section{Acknowledgments}
We thank Jeremy Blackburn, Oana Goga, Krishna Gummadi, Shagun Jhaver, and Manoel Horta Ribeiro for fruitful discussions and feedback during this work. Also, we thank the anonymous reviewers for providing useful suggestions and constructive criticism that helped improve this work.

\bibliographystyle{abbrv}
\bibliography{references}

\begin{thebibliography}{10}

\bibitem{bode2015related}
L.~Bode and E.~K. Vraga.
\newblock In related news, that was wrong: The correction of misinformation
  through related stories functionality in social media.
\newblock {\em Journal of Communication}, 65(4):619--638, 2015.

\bibitem{braun2006using}
V.~Braun and V.~Clarke.
\newblock Using thematic analysis in psychology.
\newblock {\em Qualitative research in psychology}, 3(2):77--101, 2006.

\bibitem{chandrasekharan2020quarantined}
E.~Chandrasekharan, S.~Jhaver, A.~Bruckman, and E.~Gilbert.
\newblock {Quarantined! Examining the Effects of a Community-Wide Moderation
  Intervention on Reddit}.
\newblock {\em arXiv preprint arXiv:2009.11483}, 2020.

\bibitem{chandrasekharan2017you}
E.~Chandrasekharan, U.~Pavalanathan, A.~Srinivasan, A.~Glynn, J.~Eisenstein,
  and E.~Gilbert.
\newblock You can't stay here: The efficacy of reddit's 2015 ban examined
  through hate speech.
\newblock In {\em CSCW}, 2017.

\bibitem{twitter_biased}
J.~Clayton.
\newblock {Social media: Is it really biased against US Republicans?}
\newblock \url{https://www.bbc.com/news/technology-54698186}, 2020.

\bibitem{capitol_attack}
K.~Evelyn.
\newblock {Capitol attack: the five people who died}.
\newblock
  \url{https://www.theguardian.com/us-news/2021/jan/08/capitol-attack-police-officer-five-deaths},
  2021.

\bibitem{twitter_changes}
V.~Gadde and K.~Beykpour.
\newblock {Additional steps we're taking ahead of the 2020 US Election}.
\newblock
  \url{https://blog.twitter.com/en_us/topics/company/2020/2020-election-changes.html},
  2020.

\bibitem{geeng2020social}
C.~Geeng, T.~Francisco, J.~West, and F.~Roesner.
\newblock {Social Media COVID-19 Misinformation Interventions Viewed
  Positively, But Have Limited Impact}.
\newblock In {\em arXiv preprint arXiv:2012.11055}, 2020.

\bibitem{ghosh2012cognos}
S.~Ghosh, N.~Sharma, F.~Benevenuto, N.~Ganguly, and K.~Gummadi.
\newblock Cognos: crowdsourcing search for topic experts in microblogs.
\newblock In {\em SIGIR}, 2012.

\bibitem{gillespie2018custodians}
T.~Gillespie.
\newblock {\em Custodians of the Internet: Platforms, content moderation, and
  the hidden decisions that shape social media}.
\newblock Yale University Press, 2018.

\bibitem{twitter_hate_speech2016}
J.~Guynn.
\newblock {``Massive rise'' in hate speech on Twitter during presidential
  election}.
\newblock
  \url{https://eu.usatoday.com/story/tech/news/2016/10/21/massive-rise-in-hate-speech-twitter-during-presidential-election-donald-trump/92486210/},
  2016.

\bibitem{twiter_updates}
A.~Hutchinson.
\newblock {Twitter Updates Hate Speech Policy to Include Links to ``Hateful''
  Content}.
\newblock
  \url{https://www.socialmediatoday.com/news/twitter-updates-hate-speech-policy-to-include-links-to-hateful-content/582482/},
  2020.

\bibitem{kaiser2020adapting}
B.~Kaiser, J.~Wei, E.~Lucherini, K.~Lee, J.~N. Matias, and J.~Mayer.
\newblock {Adapting Security Warnings to Counter Misinformation}.
\newblock In {\em Usenix Security}, 2020.

\bibitem{kulshrestha2017quantifying}
J.~Kulshrestha, M.~Eslami, J.~Messias, M.~B. Zafar, S.~Ghosh, K.~P. Gummadi,
  and K.~Karahalios.
\newblock Quantifying search bias: Investigating sources of bias for political
  searches in social media.
\newblock In {\em CSCW}, 2017.

\bibitem{kulshrestha2019search}
J.~Kulshrestha, M.~Eslami, J.~Messias, M.~B. Zafar, S.~Ghosh, K.~P. Gummadi,
  and K.~Karahalios.
\newblock Search bias quantification: investigating political bias in social
  media and web search.
\newblock {\em Information Retrieval Journal}, 22(1-2):188--227, 2019.

\bibitem{mena2020cleaning}
P.~Mena.
\newblock Cleaning up social media: The effect of warning labels on likelihood
  of sharing false news on facebook.
\newblock {\em Policy \& internet}, 12(2):165--183, 2020.

\bibitem{merrer2020setting}
E.~L. Merrer, B.~Morgan, and G.~Tr{\'e}dan.
\newblock Setting the record straighter on shadow banning.
\newblock {\em arXiv preprint arXiv:2012.05101}, 2020.

\bibitem{inference_api}
J.~Messias.
\newblock {Political Bias Inference API}.
\newblock \url{https://github.com/johnnatan-messias/bias_inference_api}, 2017.

\bibitem{moravec2020appealing}
P.~L. Moravec, A.~Kim, and A.~R. Dennis.
\newblock Appealing to sense and sensibility: System 1 and system 2
  interventions for fake news on social media.
\newblock {\em Information Systems Research}, 31(3):987--1006, 2020.

\bibitem{myers2018censored}
S.~Myers~West.
\newblock Censored, suspended, shadowbanned: User interpretations of content
  moderation on social media platforms.
\newblock {\em New Media \& Society}, 2018.

\bibitem{newell2016user}
E.~Newell, D.~Jurgens, H.~M. Saleem, H.~Vala, J.~Sassine, C.~Armstrong, and
  D.~Ruths.
\newblock {User Migration in Online Social Networks: A Case Study on Reddit
  During a Period of Community Unrest.}
\newblock In {\em ICWSM}, pages 279--288, 2016.

\bibitem{nyhan2010corrections}
B.~Nyhan and J.~Reifler.
\newblock When corrections fail: The persistence of political misperceptions.
\newblock {\em Political Behavior}, 2010.

\bibitem{pennycook2020implied}
G.~Pennycook, A.~Bear, E.~T. Collins, and D.~G. Rand.
\newblock The implied truth effect: Attaching warnings to a subset of fake news
  headlines increases perceived accuracy of headlines without warnings.
\newblock {\em Management Science}, 2020.

\bibitem{pennycook2018prior}
G.~Pennycook, T.~D. Cannon, and D.~G. Rand.
\newblock Prior exposure increases perceived accuracy of fake news.
\newblock {\em Journal of experimental psychology: general}, 147(12):1865,
  2018.

\bibitem{pushshift_verified}
{Pushshift}.
\newblock {Pushshift's Twitter Verified Users Dataset}.
\newblock
  \url{https://files.pushshift.io/twitter/TW_verified_users.ndjson.zst}, 2020.

\bibitem{ribeiro2020does}
M.~H. Ribeiro, S.~Jhaver, S.~Zannettou, J.~Blackburn, E.~De~Cristofaro,
  G.~Stringhini, and R.~West.
\newblock {Does Platform Migration Compromise Content Moderation? Evidence from
  r/The\_Donald and r/Incels}.
\newblock {\em arXiv preprint arXiv:2010.10397}, 2020.

\bibitem{rivers2014ethical}
C.~M. Rivers and B.~L. Lewis.
\newblock {Ethical research standards in a world of big data}.
\newblock {\em F1000Research}, 2014.

\bibitem{facebook_softmod}
G.~Rosen.
\newblock {An Update on Our Work to Keep People Informed and Limit
  Misinformation About COVID-19}.
\newblock \url{https://about.fb.com/news/2020/04/covid-19-misinfo-update/},
  2020.

\bibitem{twitter_softmod}
Y.~Roth and N.~Pickles.
\newblock {Updating our approach to misleading information}.
\newblock
  \url{https://blog.twitter.com/en_us/topics/product/2020/updating-our-approach-to-misleading-information.html
  }, 2020.

\bibitem{saleem2018aftermath}
H.~M. Saleem and D.~Ruths.
\newblock {The Aftermath of Disbanding an Online Hateful Community}.
\newblock {\em arXiv preprint arXiv:1804.07354}, 2018.

\bibitem{saltz2020encounters}
E.~Saltz, C.~Leibowicz, and C.~Wardle.
\newblock {Encounters with Visual Misinformation and Labels Across Platforms:
  An Interview and Diary Study to Inform Ecosystem Approaches to Misinformation
  Interventions}.
\newblock {\em arXiv preprint arXiv:2011.12758}, 2020.

\bibitem{seo2019trust}
H.~Seo, A.~Xiong, and D.~Lee.
\newblock {Trust It or Not: Effects of Machine-Learning Warnings in Helping
  Individuals Mitigate Misinformation}.
\newblock In {\em WebSci}, 2019.

\bibitem{sharma2012inferring}
N.~K. Sharma, S.~Ghosh, F.~Benevenuto, N.~Ganguly, and K.~Gummadi.
\newblock Inferring who-is-who in the twitter social network.
\newblock {\em SIGCOMM CCR}, 2012.

\bibitem{facebook_disinformation2016}
O.~Solon.
\newblock {Facebook's failure: did fake news and polarized politics get Trump
  elected?}
\newblock
  \url{https://www.theguardian.com/technology/2016/nov/10/facebook-fake-news-election-conspiracy-theories},
  2016.

\bibitem{swire2020searching}
B.~Swire-Thompson, J.~DeGutis, and D.~Lazer.
\newblock Searching for the backfire effect: Measurement and design
  considerations.
\newblock {\em Journal of Applied Research in Memory and Cognition}, 2020.

\bibitem{twitter_notices}
{Twitter}.
\newblock {Notices on Twitter and what they mean}.
\newblock
  \url{https://help.twitter.com/en/rules-and-policies/notices-on-twitter},
  2020.

\bibitem{twitter_trump_suspension}
{Twitter Inc.}
\newblock {Permanent suspension of \@realDonaldTrump}.
\newblock
  \url{https://blog.twitter.com/en_us/topics/company/2020/suspension.html},
  2021.

\bibitem{wood2019elusive}
T.~Wood and E.~Porter.
\newblock The elusive backfire effect: Mass attitudes’ steadfast factual
  adherence.
\newblock {\em Political Behavior}, 41(1):135--163, 2019.

\bibitem{dataset}
S.~Zannettou.
\newblock {Dataset of Soft Moderation Interventions on Twitter}.
\newblock
  \url{https://github.com/zsavvas/Soft-Moderation-Interventions-Twitter}, 2021.

\end{thebibliography}

\appendix

\section{Validation of the Political Bias Inference API} \label{sec:appendix}

Kulshrestha et al.~\cite{kulshrestha2017quantifying} measured the performance of the Political Bias Inference API; they found an accuracy of 92\% for US senators in 2017 and 85\% for Twitter users that self-identified their political views on Twitter. 
Given that our dataset also includes users that are not US senators and that do not self-identify their political views on Twitter, we evaluated the performance of the Political Bias Inference API in our dataset.
To validate the performance of the Political Bias Inference API, we extracted a sample of 200 users that were independently classified as Republican/Democrat/Neutral by the first author of this study. 
The sample includes the top 100 users in our dataset (in terms of the tweets with warning labels/quoted tweets) and another 100 randomly sampled users (we ensured no overlap between the two sets). 
The manual classification task for each user was undertaken as follows. 
First, we inspected the Twitter user profile to check if they report their political views/party. 
Second, we searched on Google to find more about the user and find references mentioning their political views.
Finally, we searched for tweets posted by the user in consideration that discussed politics to infer their political views. 
Then, we compared the manual classification with the inferences from the Political Bias Inference API; we found an average accuracy of 80.5\% (82\% for the top 100 users and 79\% of the randomly selected 100 users).
Even though the API has a considerable number of misclassifications, we believe that it can provide a meaningful signal with regards to the political views of the users in our dataset.

\end{document}